# CONSTRUCTIVE COMMENTS ON QED2 IN CURVED SPACETIME - I

T. Mark Harder and Thomas K. Delillo [†]

[†] DEPARATMENT OF MATHEMATICS & PHYSICS, WICHITA STATE UNIVERSITY, 1845 FAIRMOUNT ST., WICHITA, KS 67260.

In this paper an axiomatic formulation of the Schwinger Model in curved spacetime is considered. The mathematical approach utilized is that specified by Hollands and Wald in [1]. We will attempt to present the theory in this context, exhibiting the operator product expansions (OPEs) that define the theory in a follow-on paper.

## INTRODUCTION

In flat spacetime the massless Schwinger model is exactly solvable, and this feature carries over to curved spacetime, as has been demonstrated in many papers. The purpose of our formulation is to present a rigorous mathematical definition of the operator algebra in curved spacetime, the extended Hilbert space of the solution, and show how the theory is defined via the OPEs. We will adopt the viewpoint expressed by F. Strocchi [2], which involves only those degrees of freedom contained in the defining equations and makes reference to a Hilbert space obtained by completion of local states according to a suitable minimum topology.

The order of presentation will begin with a heuristic presentation of the model and its curved spacetime (classical) solution. This will serve to define the notation and focus the discussion on the objects whose rigorous definitions will follow. The definition and conditions of basic and auxiliary fields, needed for the exact solution, is relegated to an appendix. The total solution is made up of these fields. A discussion of the representation of charged states follows. All of this is old ground. Finally we will present a rigorous definition of the solution.

In part II of this series we will construct the consequent OPEs and show that they satisfy the axioms specified in [1]. The exposition in part I is intentionally addressed to someone not over familiar with the details of quantum fields in curved spacetime, or even some of the more detailed constructive ideas in quantum field theory.

______________________________

*E-mail address:* tmharder1@wichita.edu
*E-mail address:* wwdelillo@math.wichita.edu

# THE SCHWINGER MODEL AND CLASSICAL THEORY

The equations in Minkowski space for massless electrodynamics in two dimensions are

| Classical Equation | Quantum Equation | |
|---|---|---|
| $i\gamma^\mu \partial_\mu \psi + e\gamma^\mu A_\mu \psi = 0$ | $i\gamma^\mu \partial_\mu \psi + e\gamma^\mu (A_\mu \psi)_{ren} = 0$ | (1) |
| $\partial^\nu F_{\mu\nu} = -ej_\mu$ | $\partial^\nu F_{\mu\nu} = -ej_\mu + a_\mu$ | (2) |

We have added the "longitudinal" term to equation (2) quantum version as required to properly define the state space, and indicated a probable renormalization requirement.

Let us establish some notation. To introduce fermions in curved space-time we need to go to a locally flat space-time with which the correspondences are established via a local system of orthogonal coordinates: the zweibeins. These satisfy the relations

$$e^{\mu a}(x)\gamma_a = \gamma^\mu(x), \qquad e^{\mu a}(x)e_\mu^b(x) = \eta^{ab} \qquad e_\mu^a(x)e_\nu^a(x) = g_{\mu\nu} \qquad (3a)$$

$$e^{\mu a}(x)e_{\nu a} = \delta_\nu^\mu, \qquad e^{\mu a}(x)e_{\mu b}(x) = \delta_b^a \qquad (3b)$$

Sometimes one writes $E_a^\mu = e_a^\mu$ as the inverse of $e_\mu^a$. The indices a,b,..etc. will denote the flat space-time; the flat spaces indices are raised and lowered by the metric $\eta_{ab}$ (signature +,−) and the curved space indices are raised and lowered by $g_{\mu\nu}$. Also $\gamma^\mu = \gamma^\mu(x) \equiv e_a^\mu(x)\hat{\gamma}^a$. Here the carat $\hat{\gamma}$ refers to the flat space quantity.

Because the metric and zweibeins are coordinate dependent an ordinary derivative of a vector or tensor does not transform as a tensor in curved spacetime. Thus one defines $\nabla_\alpha A_\mu \equiv \partial_\alpha A_\mu - \Gamma_{\alpha\mu}^\nu A_\nu$, and $\nabla_\alpha A^\mu \equiv \partial_\alpha A + \Gamma_{\alpha\nu}^\mu A^\nu$. Since it is required that $\nabla_\alpha$ acting on a spinor must transform as a spinor, one must be able to pull any Dirac matrix through this derivative. This has the consequence that $\nabla_\alpha \gamma^\alpha = 0$. One then defines a spin connection $\Gamma_\mu = \frac{1}{2}\omega_{\mu ab}\Sigma^{ab}$.

With this, one arrives at the action for the Dirac equation in curved spacetime

$$S_D = \int d^D x \sqrt{-g}\, \bar{\psi}(i\gamma^\alpha e_a^\mu (\partial_\mu \psi + \frac{1}{2}\omega_{\mu bc}\Sigma^{bc}\psi) - m\psi) \qquad (4)$$

from which a variational principle gives $i\gamma^\alpha e_a^\mu(\partial_\mu \psi + \frac{1}{2}\omega_{\mu ab}\Sigma^{bc}\psi) - m\psi = 0$. Making the reduction to 1 + 1 dimensions, we select the following chiral representation for the flat space $\gamma$'s.

$$\hat{\gamma}^0 = \begin{pmatrix} 0 & 1 \\ 1 & 0 \end{pmatrix} \quad \hat{\gamma}^1 = \begin{pmatrix} 0 & 1 \\ -1 & 0 \end{pmatrix} \quad \gamma^5 = \hat{\gamma}^0 \hat{\gamma}^1 = \begin{pmatrix} -1 & 0 \\ 0 & 1 \end{pmatrix}$$

Write the gauge covariant derivative for spinors as

$$\nabla_\mu \psi = (D_\mu - ieA_\mu)\psi \text{ where } D_\mu = \partial_\mu + \tfrac{1}{2}\eta_{ac}\omega^a_{\mu b}\Sigma^{bc} \text{ and } \Sigma^{ab} = \tfrac{1}{4}[\hat{\gamma}^a, \hat{\gamma}^b]. \tag{5}$$

In two dimensions the formulae are particularly simple. It is easiest to compute the spin connection by varying the Dirac action with respect to $\delta\psi$. When integrating by parts one gets the result

$$\partial_\mu(\sqrt{-g}E^\mu_a) = \sqrt{-g}E^\mu_b \omega^b_{\mu a}. \tag{6}$$

There are some common conventions for 1 + 1 dimensional spacetimes which we will introduce here: First $\omega_{\mu ab} = -\omega_{\mu ba}$. In two dimensions the antisymmetric symbol $\varepsilon_{ab}$ is defined to be equal to $-\varepsilon^{ab}$ with $\varepsilon_{01} = +1$. The spin symbol $\omega_{\alpha ab}$ has only one unique component $\omega_{\alpha 01}$, due to antisymmetry. It is common to define a spin connection field with no Lorentz indices and write $\omega_\alpha \equiv \omega_{\alpha 01}$ and $\omega_{\alpha ab} = \varepsilon_{ab}\omega_\alpha$. Also, due to the canonical anti-commutation relations (CARs) of the $\gamma$ matrices, in 2D one can write $\Sigma^{ab} = \tfrac{1}{2}\varepsilon^{ab}\gamma^5$. The free massless Dirac equation in curved spacetime can thus be written as

$$i\hat{\gamma}^a e^\mu_a (\partial_\mu + \tfrac{1}{2}\omega_\mu \gamma^5)\psi = 0. \tag{7}$$

We deal with globally hyperbolic spacetimes, where the Cauchy problem is well posed. The manifold is generally of the form $M = \mathbb{R} \otimes M$ with the time variable being orientable and globally identifiable as positive or negative. The metric is thus of the form $ds^2 = dt^2 - h(x,t)dx^2$; with $h$ the induced metric on the foliated Cauchy surfaces $\Sigma(x,t) \in M$.

All 2 dimensional spacetimes are conformally flat, which means there exist coordinate systems in which the metric takes the form $g_{\mu\nu}(x) = \Omega(x)^2 \eta_{\mu\nu}$. Such coordinate systems do not necessarily cover the entire manifold. One can always construct such a coordinate system if the operator $\nabla^\alpha \nabla_\alpha$ is invertible. The metric then becomes $ds^2 = g(dx_0^2 - dx_1^2)$ for which one writes

$$ds^2 = e^{2\sigma}(dx_0^2 - dx_1^2); \text{ e.g. isothermal coordinates.} \tag{8}$$

A short calculation yields, $\omega^{01}_\mu = \omega_{\mu 01} \to \omega_{001} = -\dfrac{\partial \sigma}{\partial x^1}; \omega_{101} = -\dfrac{\partial \sigma}{\partial x^0};$ and $\omega_{\mu 10} = -\omega_{\mu 01}$

When applied to equation (7) these relations yield (with $\hat{\partial} = \hat{\gamma}^\mu \partial_\mu$ )

$$ie^{-\sigma}\hat{\partial}\psi + ie^{-\sigma}\tfrac{1}{2}(\hat{\partial}\sigma)\psi = 0.  \qquad (9)$$

This can also be written $i\mathbb{D}_0\psi = 0$ where $\mathbb{D}_0 = e^{-\frac{3}{2}\sigma}\hat{\partial}e^{+\frac{1}{2}\sigma}$. The solution is then written as

$\psi(x) \equiv e^{-\frac{1}{2}\sigma}\psi_0(x)$; which solves (9) when $\psi_0$ solves the flat space Dirac equation. [By this we mean that $\psi_0$ has the same functional form as the flat space solution in Minkowski variables labeled *x*.] Further along these lines, if the electromagnetic field is included, equation (9) will become

$$i\hat{\gamma}^a e_a^\mu (\partial_\mu + \tfrac{1}{2}\omega_\mu \gamma^5 - ieA_\mu)\psi = 0 \ . \qquad (10)$$

Then one still writes $i\mathbb{D}\psi = 0$, where now $\mathbb{D} = e^{iF - i\gamma^5 G - \frac{3}{2}\sigma}\hat{\partial}e^{-iF - i\gamma^5 G + \frac{1}{2}\sigma}$; $F = e\alpha$, $G = e\varphi$; and here the gauge field has been decomposed as $A_\mu = \partial_\mu \alpha - \eta_{\mu\rho}\partial^\rho \varphi$ so that $F_{01} = \sqrt{-g}\nabla^2 \varphi$

and the solution is written as $$\psi = e^{iF + i\gamma^5 G - \frac{1}{2}\sigma}\psi_0(x) \ . \qquad (11)$$

The relation $\eta_{\mu\nu} = \sqrt{-g}\,\varepsilon_{\mu\nu} = e^{2\sigma}\varepsilon_{\mu\nu}$ has been used. Again, $\psi_0(x)$ solves the free Dirac equation in flat spacetime. These exponential manipulations are common, and can found in [3]. A very nice presentation on the spinor connection which is used here is in [4]. Pursuing this classical commentary, the vector

currents are $$j^\mu = \bar{\psi}\gamma^\mu \psi = \bar{\psi}_0 \hat{\gamma}^\mu \psi_0 e^{-2\sigma} \equiv \frac{1}{\sqrt{-g}} j_0^\mu . \qquad (12)$$

The same holds for the axial vector current $\bar{\psi}\gamma^\mu \gamma_5 \psi$. $\sqrt{-g}\nabla_\mu j^\mu = \partial_\mu \sqrt{-g}\, j^\mu$ implies the conservation of the (free) vector and axial currents $\nabla_\mu j^\mu = \nabla_\mu j^{5\mu} = 0$. These conservation laws imply that the currents are free fields $$\nabla^2 j^\mu = \nabla^2 j^{5\mu} = 0 \ , \qquad (13)$$

which accounts for the solubility of the model. This also simplifies the electromagnetic field equation as $D_\nu F^{\mu\nu} = ej^\mu$; the field then being the result of a source that is a free field. It should be remarked that in curved spacetime the conjugate spinor $\bar{\psi} = \psi^* \hat{\gamma}^0$ is usually defined with the flat space gamma matrix $\hat{\gamma}^0$. The $\hat{\gamma}^5 = \gamma^5$ matrix maintains its form in general spacetimes. Of course equation (13) does not survive quantization intact.

## DISCUSSION OF CHARGED STATES

In the quantum theory of the Schwinger model in *Minkowski spacetime*, there are many manipulations that have become common to the model but may appear mysterious to the unfamiliar reader. Schwinger's original papers [5] were devoted to the issue of the existence of massive bosons in a gauge invariant theory and were presented entirely in terms of Green functions and spectral theory. The standard formalism was introduced by Lowenstein and Swieca [6], who introduced the operator algebra, auxiliary fields, and an exact operator solution from which one could derive properties and prove theorems. These all have to do with known difficulties in defining a consistent quantum field theory in 1 + 1 dimensions: (1) The interpretation of Gauss' law in a local theory and the existence of charged states; (2) The proper definition of a massless 2D scalar field that handles infrared divergences; (3) The utilization of an indefinite metric and a suitable topological completion to allow definition of an appropriate Hilbert space; (4) The construction of correlation functions with the requisite properties.

With respect to issue (1) consider the following. An element $A$ of a quantum field Algebra carries a charge $q$ if

$$[Q, A] = qA \text{ where } Q \text{ is the electric charge.} \tag{14}$$

This is defined as a limit of a local charge $Q_\rho$ where

$$Q_\rho = \int j_0(\mathbf{x}, t) f_\rho(\mathbf{x}) \alpha(t) d\mathbf{x} dt \tag{15}$$

and the following conditions hold:

$$f_\rho(\mathbf{x}) = f(|\mathbf{x}|/\rho), f \in D(\mathbb{R}), \text{ with } \quad f(\mathbf{x}) = 1 \text{ for } |\mathbf{x}| < 1, \tag{16}$$

$$f(x) = 0 \text{ for } |\mathbf{x}| > 1 + \varepsilon, \ \int \alpha(t) dt = 1, \alpha \in \mathbb{R}.$$

One then writes $\lim_{\rho \to \infty}[Q_\rho, A] = qA$. Now here the current $j_0$ is a component of a current vector that is the divergence of a local antisymmetric tensor: $j_\mu = \partial^\nu F_{\mu\nu}$, e.g. a local Gauss' law. These definitions have two results: (1) a field carrying a non-zero charge, whose current obeys a local Gauss law cannot be local relative to $F_{\mu\nu}$. (2) In a local formulation, all physical states have zero charge; (see [2]). To achieve charged states, the algebra must be enlarged.

Let us consider the following. In Minkowski space of 1+1 dimensions a solution to the massless free Klein Gordon equation can be written

$$\varphi(x, t) = \varphi_R(x - t) + \varphi_L(x + t). \tag{17}$$

From these, one constructs the dual

$$\tilde{\varphi}(x, t) = \varphi_R(x - t) - \varphi_L(x + t). \tag{18}$$

Since this satisfies $\frac{\partial \tilde{\varphi}}{\partial x} = -\frac{\partial \varphi}{\partial t}$, the momentum canonical to $\varphi(x,t)$ can be written

$$\frac{\partial \tilde{\varphi}}{\partial x} = -\pi. \tag{19}$$

Reversing the transformation one obtains $\varphi_R(x-t) = \frac{1}{2}(\varphi + \tilde{\varphi})$ and $\varphi_L(x-t) = \frac{1}{2}(\varphi - \tilde{\varphi})$. We have seen that $\varepsilon_{\mu\nu}\partial^\nu \varphi = \partial_\mu \tilde{\varphi}$. We say that $\tilde{\varphi}$ is charged with respect to a charge $\tilde{Q}_R \equiv (\partial_0 \tilde{\varphi})(f_R)$ which obeys a local Gauss law [Minkowski space notation]

$$j_\mu \equiv \partial_\mu \tilde{\varphi} = \partial^\nu \varepsilon_{\mu\nu}\varphi \equiv \partial^\nu F_{\nu\mu}. \tag{20}$$

It cannot therefor be local with respect to $\varphi$. One takes the position that the field $\varphi$ has been enlarged by the introduction of the independent fields $\varphi_R$ and $\varphi_L$ which both satisfy the free wave equation and such that $\partial_+ \varphi_R = 0$, $\partial_- \varphi_L = 0$ where $\partial_\pm = \partial/\partial_\pm$, and $x_\pm = x_0 \pm x_1$. We have introduced the *operators* necessary to extend the algebra to cover charge. The charge *states* themselves require the enlargement of the state space by the introduction of a *Krein* structure.

A standard result from the study of gauge theories are the difficulties associated with the simultaneous requirements of positivity of Wightman functions and locality. Positivity of the two-point functions is required to be able to obtain physical states by the application of smeared fields to the vacuum, along with positivity of the norm of the underlying Hilbert space. The solution requires the weakening of the local Gauss law. The local Gauss law is modified by the introduction of a "longitudinal" field $a_\mu$ to the electromagnetic field equation to yield $j_\mu = \partial^\nu F_{\nu\mu} + a_\mu$. The Hilbert space of the theory is then allowed to contain unphysical states, compensated by requiring $a_\mu$ have zero matrix elements between physical states.

A massless scalar field in 2 (Minkowski) dimensions is defined as the solution of $\partial^\mu \partial_\mu \varphi(x) = 0$ satisfying the commutation relations. If one seeks the most general two point function as solution of the wave equation consistent with Poincaré invariance one gets [7]

$$\omega_\varphi(x) = -(4\pi)^{-1}\ln(-x^2 + i\varepsilon x_0) + c, \tag{21}$$

where $c$ is fixed so that $\langle \varphi(x) \rangle = 0$. The Fourier transform of this fails to satisfy positivity, and the Wightman functions will not provide a Hilbert topology. In general (for Minkowski space), physically allowable states break Lorentz invariance. It may also be shown that in any state free of infrared divergences, $\langle \varphi^2 \rangle$ must be a growing function of time. This situation persists in curved spacetime (see [8]) and similar to the Minkowski case demands the introduction of a

*Krein* topology, which we will now define. We adapt the methods cited in the reference to the present case.

First observe that the Cauchy problem for the scalar field $\varphi(x)$ in 2D is well posed. The field can be quantized by the method of Wald [9] (appendix A) and the field algebra generated by the method of Dimock [10]. This yields a local algebra for which charged states cannot exist and one must introduce the Krein structure. The Krein construction ensures that the Hilbert space obtained by closing the space of local states is a "maximal" space. Essentially this insures that one may associate the so obtained Hilbert space to a set of Wightman functions. This association is local, and not necessarily unique with regard to large distance behavior of the class of states one can associate to the Wightman functions [2].

To begin the Krein construction in curved spacetime, let the test function space be $\mathcal{D}(X) = \{f \in C_0^\infty(M) \mid f \in L^2(M)\}$ for our 2D manifold $M$. The requirements of (1) positivity and (2) locality are defined by

(1) $\quad \langle f, f \rangle = \int_{X \otimes X} \omega_\varphi(x,y) f(x) \overline{f}(y) d\hat{\sigma}(x) d\hat{\sigma}(y) \geq 0 \quad \forall f \in \mathcal{D}(X)$ for a measure $\hat{\sigma}(x)$ on $M$

(2) $\quad \omega_\varphi(x,y) = \omega_\varphi(y,x)$ for space-like separated $x,y$

The invariant two-point function $\omega_\varphi$ is positive on the subspace of null integral test functions $\mathcal{D}_0(X) = \{f \in \mathcal{D}(X) \mid \int_X f(x) d\hat{\sigma}(x) = 0\}$. Choose a real test function $h \in \mathcal{D}(X)$ such that

$$\int_X h(x) d\hat{\sigma}(x) = 1, \quad \langle h, h \rangle = 0 \quad \text{where } \langle \cdot, \cdot \rangle \text{ is the inner product over } H_h \text{ (the one dimensional space generated by } h\text{)}.$$

Then for any $f \in \mathcal{D}(X)$ we can decompose $f$ as

$$f(x) = f_0(x) + \left( \int_X f(y) d\hat{\sigma}(y) \right) h(x) \text{ (where } f_0 \in \mathcal{D}_0\text{)}.$$

This has the consequence that the test function space $\mathcal{D}(X)$ can be decomposed as the direct sum $\mathcal{D}(X) = \mathcal{D}_0 + H_h$. The decomposition yields an indefinite sesquilinear form given by

$$\langle f, g \rangle = \langle f_0, g_0 \rangle + \int_X \overline{f}(x) d\hat{\sigma}(x) \langle h, g \rangle + \int_X g(y) d\hat{\sigma}(y) \langle f, h \rangle \quad . \tag{22}$$

Let us define an inner product $(\cdot, \cdot)$ on $\mathcal{D}(X)$ by

$$(f, g) = \langle f_0, g_0 \rangle + \langle f, h \rangle \langle h, g \rangle + \int_X \overline{f}(x) d\hat{\sigma}(x) \int_X g(y) d\hat{\sigma}(y) \equiv \langle f, \eta^1 g \rangle. \tag{23}$$

Considered as a semi-norm $(f, f)$, equation (23) clearly dominates (22) because for any complex $a, b$, $|a|^2 + |b|^2 \geq |\overline{a}b + \overline{b}a|$. This inner product is positive semi-definite and hence defines a pre-Hilbert space. Now consider the subspace $I_h = \{f \in \mathcal{D}_0 \mid \langle f_0, f_0 \rangle = 0, \langle h, f \rangle = 0\}$. A Hilbert

space is obtained by quotienting $\mathcal{D}(X)$ with respect to $I_h$ and completing in the topology defined by (23). Denote this $K$. The complete Hilbert space would then be $K = \bar{H}_h = \sum_n \otimes^n K^{(1)}$. We denote by $K^{(1)}$ the "single particle space" in a quantum theory.

We state without rigorous proof, two observations on the above:
(1) The linear functional on $K$ defined by $F_h(f) = \langle h, f \rangle$ has norm equal to one and therefore it defines a normalized element $v_0$ of $K^{(1)}$ such that $(v_0, f) = F_h(f)$. Furthermore for all $f \in \mathcal{D}(X)$,
$$\langle v_0, f \rangle = \int_X f(x) d\hat{\sigma}(x) \tag{24}$$

(2) The Hilbert space $K$ is a Krein space and can be written as a direct sum
$$K = \overline{\left(\mathcal{D}_0^\perp / I_h\right)}^{\langle \cdot, \cdot \rangle} \oplus V_0 \oplus H_h \tag{25}$$

where $\mathcal{D}_0^\perp$ is the subspace of $\mathcal{D}_0$ orthogonal to $v_0$ spanning $V_0 = (\{\lambda v_0 : \lambda \in \mathbb{C}\} + I_h)$ and $H_h = (\{\lambda h : \lambda \in \mathbb{C}\} + I_h) / I_h$. The metric operator $\eta^{(1)}$ defined by $\langle \cdot, \cdot \rangle = (\cdot, \eta^{(1)} \cdot)$ is given by
$$\eta^{(1)}_{\overline{\left(\mathcal{D}_0^\perp / I_h\right)}^{\langle \cdot, \cdot \rangle}} = \mathbf{1}_{\overline{\left(\mathcal{D}_0^\perp / I_h\right)}^{\langle \cdot, \cdot \rangle}}, \quad \eta^{(1)} h = v_0, \quad \eta^{(1)} v_0 = h \tag{26}$$

<u>Sketch of proof</u>: (Taken from reference [8]) Since for any $f \in \mathcal{D}_0$ we have $(f, v_0) \propto \langle f, h \rangle$ it is true that $I_h$ is completely contained in $\mathcal{D}_0^\perp$. Separate the space $\mathcal{D}_0$ as $\mathcal{D}_0 = \mathcal{D}_0^\perp + \mathbb{C}\{v_0\}$ where the orthogonality is with respect to the pre-Hilbert product $(\cdot, \cdot)$. On the space $\mathcal{D}_0^\perp$ the Krein product is equal to the indefinite form $\langle \cdot, \cdot \rangle$. For any pair of test functions $f, g \in \mathcal{D}_0^\perp$, one has $(f, g) = \langle f, g \rangle + \langle f, h \rangle \langle h, g \rangle = \langle f, g \rangle + \langle f, v_0 \rangle \langle v_0, g \rangle = \langle f, g \rangle$. Now $H_h$ is orthogonal to $V_0$ because $(v_0, h) = \langle h, h \rangle = 0$. Additionally $\mathcal{D}_0^\perp \perp H_h$ because $(h, f) = \int_X d\sigma(x) h(x) \int_X d\sigma(y) f(y) = 0$. This is true also for $\overline{\mathcal{D}_0^\perp}^{\langle \cdot, \cdot \rangle}$, proving the decomposition (25). The definition of the metric is obvious.

<u>Comment on (1)</u>: In contrast to the indicated reference the function $v_0$ may be explicitly constructed as an appropriate limit that belongs only to the Krein completion of the test function space. Suppose $I_h$ is the ideal of the test functions of zero norm as defined above. We can define $K = \overline{\left(\mathcal{D}_0^\perp / I_h\right)}^{\|\cdot\|}$. Pick $v_0 \in I_h$ and then construct a sequence $v_n$ such that $v_n$ converges to zero on

$\left(\overline{\mathcal{D}_0^\perp / I_h}\right)^{\langle\cdot,\cdot\rangle}$, and keeps finite the norm $(v_n, v_n) = 1$. Then the desired properties for the Krein space are obtained: $\langle v_0, g \rangle = 0, \forall g \in \left(\overline{\mathcal{D}_0^\perp / I_h}\right)^{\langle\cdot,\cdot\rangle}$, $\langle v_0, v_0 \rangle = 0$ and $(v_0, v_0) = 1$.

In order to see what all this has to do with charge, let us investigate the "gauge" automorphism $\gamma^\lambda : \varphi(x) \to \varphi(x) + \lambda c$. In this case let $c = \langle f, h \rangle$. This argument is a curved space adaptation of reference [21].

The massless scalar field can be looked at as producing a conserved vector field via $\varphi^v_{;v} = \frac{1}{\sqrt{-g}} \frac{\partial}{\partial x^v}(\sqrt{-g}\varphi^v) = 0$, and thus $\frac{1}{\sqrt{-g}} \frac{\partial}{\partial x^v} \varphi$ is conserved. Consider the field algebra $\mathcal{F}$ generated by the fields $\varphi(f)$ and the Wick exponential $:e^{z\varphi}:(f)$. Using the Krein construction, decompose $\varphi(v_0)$ by $\varphi(v_0) = \varphi_+(v_0) + \varphi_-(v_0)$, where the splitting is the positive and negative energy parts. These satisfy the commutation relations $[\varphi(v_0), \varphi(f)] = 0$, and $[\varphi_-(v_0), \varphi(f)] = \langle [\varphi_-(v_0), \varphi(f)] \rangle = \langle v_0, f \rangle = \langle h, f \rangle$.

## OPERATOR SOLUTION

Consider a massive scalar field $\Sigma$ over a two dimensional globally hyperbolic spacetime $M$. Let us also define $\eta_{\mu\nu} \partial^\nu \Sigma \equiv \sqrt{-g} \varepsilon_{\mu\nu} \partial^\nu \Sigma = \tilde{\partial}_\mu \Sigma$. Define $\mathcal{A}(\Sigma)$ as the algebra generated by $\Sigma$. Introduce a massless scalar field $\eta$ and its dual similarly defined, but require that its two point function satisfy

$$\omega_\eta^+(x, y) = -\omega_\varphi(x, y) \text{ as the latter is defined in equation (21)} \tag{27}$$

Define $\mathcal{A}(\eta)$ as its algebra. To satisfy equation (27) one must write the inner product for $\eta$ as

$(f, g) = -\langle f_0, g_0 \rangle + \langle f, h \rangle \langle h, g \rangle + \int_X \overline{f}(x) d\hat{\sigma}(x) \int_X g(y) d\hat{\sigma}(y)$ to insure a positive definite form for $\eta$ on its associated Krein space, with $h$ defined as above: $\int_X h(x) d\sigma(x) = 0$, $\langle h, h \rangle = 0$. With duals appropriately defined, we may consider the enlarged algebras $\mathcal{A}(\tilde{\Sigma})$ and $\mathcal{A}(\tilde{\eta})$. Finally, as above, consider also a free, massless Dirac field $\psi_0$ and its associated algebra $\mathcal{A}(\overline{\psi}_0, \psi_0)$. We will assume these algebras have been defined in terms of local, covariant quantum fields.

For the exact solution to the Schwinger model, it is necessary to expand the algebras for $\mathcal{A}(\tilde{\Sigma})$ and $\mathcal{A}(\tilde{\eta})$ to include Wick polynomials and Wick powers of arbitrary order. The procedure for doing this is explained in [16]. This requires attaching a meaning to expressions of the form $:\varphi(x_1)\varphi(x_2)...\varphi(x_n):$ and $:\varphi(x)^n:$ that corresponds to standard usage in Minkowski space. For the first of these one defines the objects

$$W_n(x_1,...,x_n) = :\varphi(x_1)...\varphi(x_n):_\omega \equiv \frac{\delta^n}{i^n \delta f(x_1)...\delta f(x_n)} \exp[\tfrac{1}{2}\omega(f \otimes f) + i\varphi(f)]|_{f=0} \quad (28)$$

where $\omega(f \otimes f)$ is the two point function. Wick's theorem is obtained in the following way: the operators $W_n(t)$ obtained by smearing $W_n(x_1,...,x_n)$ with $t = f_1 \otimes \cdots \otimes f_n \in \mathcal{D}(M^n)$ are in the algebra $\mathcal{A}(\varphi)$, and the product of two such operators $W_n(t)$ and $W_m(t')$ is given by

$$W_n(t)W_m(t') = \sum_k W_{n+m}(t \otimes_k t') \quad \forall t \in \mathcal{D}(M^n), t' \in \mathcal{D}(M^m). \quad (29)$$

Now
$$t \otimes_k t' = \mathbf{S} \frac{n!m!}{(n-k)!(m-k)!k!} \int_{M^{2k}} t(y_1,...,y_k,x_1,...,x_{n-k}) \times$$

$$t'(y_{k+1},...,y_{k+i},x_{n-k+1},...,x_{n+m-2k}) \prod_{i=1}^k \omega(y_i, y_{k+i}) \mu_g(y_i) \mu_g(y_{k+i}) \quad (30)$$

is the symmetrized ($\mathbf{S}$ means symmetrization in $x_1$ to $x_{n+m-2k}$), $k$ times contracted tensor product. If $n < k$ or $m < k$, then the contracted tensor product is defined to be zero.

By taking $t(x_1,...,x_k) = f(x_1)\delta(x_1,...x_k)$ one arrives at $W_k(t) =: \varphi^k(f):$ provided the distribution $t$ is in a wave front set comprised of suitably defined test functions. We will assume that this is the case. An additional issue with these constructions is locality. Following reference [17], we modify equation (29) to the form (and defining the normal product in lieu of $W_n$)

$$:\varphi(x_1)...\varphi(x_n):_H \equiv \frac{\delta^n}{i^n \delta f(x_1)...\delta f(x_n)} \exp[i\varphi(f) + \tfrac{1}{2}H(f,f)\cdot\mathbf{1}]|_{f=0}. \quad (31)$$

Here $H$ is the Hadamard parametrix $H(x,y) = V(x,y)\ln\bar{\sigma}$, particular to the field (but not the state) suitably smeared. This procedure makes sense only in a sufficiently small neighborhood of the diagonal. Utilizing equation (31) and the smearing functions (30), a functional derivative of $W_k(t) =: \varphi^k(f):_H$ will yield $:\varphi^k(x):_H$, the $k^{\text{th}}$ order Wick polynomial in $x$, of Hadamard form. Now one may define a Wick exponential distribution by (see [7]) the formal expression

$$:e^{i\alpha\varphi}:(x) = \sum_{n=0}^\infty \frac{\alpha^n}{n!} :\varphi^n:(x) \quad (32)$$

Smeared with a test function in $\mathcal{D}(X) = \{f \in C_0^\infty(M) \mid f \in L^2(M)\}$, (32) makes sense on the vacuum $\Psi_0$ because $\left\| \sum_{n=0}^{\infty} \frac{\alpha^n}{n!} :\varphi^n:(f)\Psi_0 \right\|^2 \leq \iint \bar{f}(x)\exp\left(|\alpha|^2 \frac{1}{i}\omega^+(x,y)\right) f(y) dx dy < \infty$.

Similarly $:\exp(\alpha_1\varphi):(f_1)...:\exp(\alpha_n\varphi):(f_n)\Psi_0$ may be defined by its power series, for $\omega^+(x,y) = \langle \Psi_0, \varphi(x)\varphi(y)\Psi_0 \rangle$. Finally, $:e^{i\alpha\varphi}:(x)$ is a legitimate field with two point function $\left(\Psi_0 : e^{i\alpha_1\varphi}:(x_1):e^{i\alpha_2\varphi}:(x_2)\Psi_0\right) = \exp\left(\frac{\alpha_1\alpha_2}{i}\omega^+(x_1,x_2)\right)$. Hereafter, we will use the heuristic convention $\langle \phi\phi \rangle \approx \frac{1}{i}\Delta^+ \approx H$.

Now the procedure is: introduce the relevant field algebra of smeared operators; select a state $\omega$ and use the GNS construction to produce a Hilbert space; introduce the appropriate Hadamard state for $\omega$ and utilize (31) for calculating normal products. In addition to the above references, [9] discusses the viability of these procedures in terms of general covariance, and dependence of the exact form for $H$ on the arbitrary state selected to initiate the GNS procedure.

In curved spacetime the quantum Schwinger system is

$$i\gamma^\mu \nabla_\mu \psi + e\gamma^\mu (A_\mu \psi)_{ren} = 0, \qquad F^{\mu\nu}{}_{;\nu} = \frac{1}{\sqrt{-g}} \frac{\partial}{\partial x^\nu}(\sqrt{-g} F^{\mu\nu}) = -eJ^\mu + a^\mu. \quad (33)$$

Here $\gamma^\mu = \gamma^\mu(x) = e_a^\mu(x)\hat{\gamma}^a$ and $\nabla_\mu = \partial_\mu + \frac{1}{2}\eta_{ac}\omega_{\mu b}^a \Sigma^{bc}$ with $\Sigma^{ab} = \frac{1}{4}[\hat{\gamma}^a, \hat{\gamma}^b]$.

Let us define 
$$(A_\mu \psi)_{ren} = \lim_{\substack{\varepsilon \to 0 \\ \varepsilon^2 < 0}} [A_\mu(x+\varepsilon)\psi(x) + \psi(x) A_\mu(x-\varepsilon)]/2. \quad (34)$$

Take $A_\mu(x) = -\frac{\sqrt{\pi}}{e}(\tilde{\partial}_\mu \Sigma(x) + \partial_\mu \tilde{\eta}(x))$ as defining the vector potential. We are assuming the Lorentz gauge; the field $\tilde{\eta}$ is a free massless field and our notation has indicated that it is a "dual" in some nature that has been completed according to a Krein procedure. If one defines $\tilde{\partial}_\mu \eta \equiv \partial_\mu \tilde{\eta}$ then we can write 
$$A_\mu(x) = -\frac{\sqrt{\pi}}{e}(\tilde{\partial}_\mu \Sigma(x) + \tilde{\partial}_\mu \eta(x)). \quad (35)$$

This results in $F_{\mu\nu} = -\frac{\sqrt{\pi}}{e}(\partial_\mu \tilde{\partial}_\nu - \partial_\nu \tilde{\partial}_\mu)\Sigma = -\frac{\sqrt{\pi}}{e}\varepsilon_{\mu\nu}\sqrt{-g}\Delta\Sigma$. From equation (11) we have

then 
$$\psi =: e^{i\sqrt{\pi}\gamma^5(\Sigma(x)+\eta(x)) - \frac{1}{2}\sigma} :\psi_0(x) \quad (36)$$

with $\psi_0(x)$ a free, zero mass "spinor" field, and normal ordering circa (32) has been indicated.

It is useful to use operator algebra to investigate the nature of the various fields we have defined. We will use the arguments of reference [8], translated to curved spacetime. Let us start with the expression $\sqrt{-g}\bar{\psi}\gamma^\mu\psi = \sqrt{-g}J^\mu = -\frac{1}{\sqrt{\pi}}\tilde{\partial}^\mu\phi$ which we will use to define $\phi$. Using the identity $\gamma^\mu\eta_{\mu\nu} = \gamma^5\gamma_\nu$ one arrives at $J_\mu^5 = \bar{\psi}\gamma_\mu\gamma^5\psi = -\frac{1}{\sqrt{\pi}}\partial_\mu\phi$. (Note $\sqrt{-g}$ is not present)

In curved spacetime the axial anomaly is $\nabla_\mu(J_5^\mu) = -\frac{1}{\sqrt{-g}}\frac{e}{2\pi}\varepsilon^{\mu\nu}F_{\mu\nu}$, see [18]; which leads to the expression $\Delta\Sigma = \Delta\phi$ or $\Sigma + h = \phi$ with $\Delta h = 0$. The current then becomes

$$\sqrt{-g}J_\mu = -\frac{1}{\sqrt{\pi}}\tilde{\partial}_\mu\Sigma + L_\mu \qquad (37)$$

with $L_\mu = -\frac{1}{\sqrt{\pi}}\tilde{\partial}_\mu h$. Setting aside the $a^\mu$ term for the moment, equation (33) leads to

$\tilde{\partial}^\nu(\Delta + \frac{e^2}{\pi})\Sigma - \frac{e^2}{\sqrt{\pi}}L^\nu = 0$. Since $\Delta h = 0$ we will have $\tilde{\partial}_\mu L_\mu = 0$ or $\Delta(\Delta + \frac{e^2}{\pi})\Sigma = 0$. If one also requires $(\Delta + \frac{e^2}{\pi})\Sigma = 0$, then the gauge field $A_\mu$ becomes a massive field. The gauge field thus has a mass of $m = \sqrt{\frac{e^2}{\pi}}$. As an operator equation, (37) would imply that $L_\mu = 0$ strongly; however we will use the existence of $a_\mu$ to reduce the requirement on $L_\mu$ to vanish only in the weak sense $\langle\psi'|L_\mu(x)|\psi\rangle = 0$. This will become a condition on the physical Hilbert space $H_{phys}$.

Define the "observables" as those operators $\mathcal{O}$ which leave $H_{phys}$ invariant: $O|\psi\rangle \in H_{phys}$ if $|\psi\rangle \in H_{phys}$. The condition on $L_\mu$ becomes $\langle\psi'|L_\mu(x)\mathcal{O}|\psi\rangle = 0$. Since the current $J_\mu$ and the field $F_{\mu\nu}$ are observables

$$\langle\psi'|L_\mu(x)L_\nu(y)|\psi\rangle = 0 \qquad (38)$$

and thus $\|L[f]|\psi\rangle\| = 0$, where $L[f] = \int f_\mu(x) L^\mu(x) \sqrt{-g} d^2x$. Define the "vacuum" by the conditions $L_\mu^{(-)}(x)|\Omega\rangle = 0$, and $\Sigma^{(-)}(x)|\Omega\rangle = 0$. Now $|\Omega\rangle$ belongs to $H_{phys}$. Chose $|\psi'\rangle = |\psi\rangle = |\Omega\rangle$ in equation (38) and one arrives at

$$\langle \Omega | \left[ L_\mu^{(-)}(x), L_\nu^{(+)}(y) \right] | \Omega \rangle = 0 \text{ , for all } x,y. \tag{39}$$

This equation has very important consequences for the Schwinger model.

We will make preliminary use of one OPE to define the operator current. Consider

$$\bar{\psi}(x+\varepsilon)\gamma^\mu : e^{-ie\int_x^{x+\varepsilon} A_\mu(\xi)d\xi^\mu} : \psi(x) = C^{(0)}(x) + Z(\varepsilon) : \bar{\psi}(x+\varepsilon)\gamma^\mu \psi(x): + R(\varepsilon) \tag{40}$$

where $R \to 0$ as $\varepsilon \to 0$. $\varepsilon$ is along a geodesic through $x$, and $C$ and $Z$ are (possibly singular) C-number functions of the parameter $\varepsilon$. In a small region around the point x, the singular behavior of the current is $J^\mu \approx \lim_{\varepsilon \to 0} -(\frac{i}{2\pi})\frac{e^{-\sigma}}{\bar{\sigma}}\partial^\mu(\bar{\sigma}) \to -\frac{i}{\pi}\frac{\varepsilon^\mu}{\varepsilon^2}$, for $2\bar{\sigma} = \varepsilon^2$ along a geodesic through the point $x$. Let us define a "potential" $\varphi$ for the free current $\sqrt{-g} : \bar{\psi}_f \gamma^\mu \psi_f : (x) = -\frac{1}{\sqrt{\pi}}\tilde{\partial}^\mu \varphi$, which implies that $:\bar{\psi}_f \gamma^\mu \gamma^5 \psi_f : (x) = -\frac{1}{\sqrt{\pi}}\partial^\mu \varphi$. We can make the identification $C^{(0)}(x) = -\frac{i}{\pi}Z(\varepsilon)\frac{\varepsilon^\mu}{\varepsilon^2}$ which leaves the following for the right hand side of equation (40), RHS $= Z(\varepsilon) : \bar{\psi}(x+\varepsilon)\gamma^\mu \psi(x): \to Z(\varepsilon)\{-\frac{1}{\sqrt{\pi}}\tilde{\partial}^\mu \varphi + \frac{e}{\pi}A^\mu(x)\}$. Now inserting the expression for $A^\mu$ gives us $\sqrt{-g} : \bar{\psi}(x)\gamma^\mu \psi(x) := -\frac{1}{\sqrt{\pi}}\tilde{\partial}^\mu [\Sigma + (\eta + \varphi)]$. So $h = \eta + \varphi$. This relation, along with equation (39) requires that the sign of the two point function of $\eta$ be opposite that normally associated with a free scalar field $\varphi$, and also with $\Sigma$. This also insures that the normalization constant $Z(\varepsilon)$ is finite. The "longitudinal field" $a_\mu$ in equation (33) is seen to be $a_\mu = eJ_\mu^L = eJ_\mu^{free} - \frac{e}{\sqrt{\pi}}\tilde{\partial}_\mu \eta(x)$; with $a_\mu \Psi = 0$ for $\Psi \in H_{phys}$.

The physical space of the problem is defined as follows. Label as $\mathcal{F}$ the algebra generated by the charged spinor field $\psi$ and by the vector potential $A_\mu$. Label as $K$ the Hilbert space of states obtained by completion of local states according to a suitable minimal topology. One then looks for solutions to the subsidiary condition $(J_\mu^L(x))^-\Omega = 0$ in $K$. We now seek to quantify the objects $\mathcal{F}$ and $K$.

Let us begin by considering the object $\psi'(x) =: e^{i\sqrt{\pi}\gamma^5\eta} :(x) e^{-\frac{1}{2}\sigma}\psi_0(x)$. Consider the semi-norms $p_\eta(f)^2 \equiv -\langle f_0, f_0 \rangle + \langle f, h \rangle\langle h, f \rangle + \int_X \bar{f}(x)d\hat{\sigma}(x) \int_X f(y)d\hat{\sigma}(y)$ with $f \in \mathcal{D}(X)$ and using the same conventions as following equation (36). This defines a Krein topology for the correlation functions of the massless scalar field $\eta$, and therefore of its Wick exponential. According to the above discussion this semi-norm, plus the semi-norms defined by the positive two point functions of the free spinor field $\psi_0$, and the scalar field $\Sigma$ define a Hilbert majorant (see [2]) topology $\tau_1$ for the correlation functions of $\mathcal{F}$. Let $H_\Sigma$ be the Fock space of the free massive field $\Sigma$. Let $\mathcal{F}'$ be the field algebra generated by $J^{\mu\,free}, J_\mu^L$ and

$\psi'(x) =: e^{i\sqrt{\pi}\gamma^5\eta} :(x) e^{-\frac{1}{2}\sigma}\psi_0(x)$. Define $H_{\eta,\psi_0} = \overline{\mathcal{F}'\Omega_{\psi'}}^\tau$ as the Hilbert space of the completion of $\mathcal{F}'$ under the Krein topology. With all this structure, it is now possible to find a Krein topology $\tau$ which majorizes the correlation functions of $\mathcal{F}$ and defines a maximal Hilbert space structure $(K, \eta_m)$ with the metric $\eta_m^2 = 1$ and $K = \overline{\mathcal{F}\Omega}^\tau$. The final result is $K = H_\Sigma \otimes H_{\eta,\psi_0}$. [See appendix D for nomenclature]. Now the physical space $K'$ is defined by the solutions of the subsidiary condition $(J_\mu^L(x))^- \Omega = 0$ in $K$.

There is one final point which is extremely important for the physical content of the Schwinger model. Is it possible to decompose $H_{\eta,\psi_0}$ as a tensor product of the Krein space of $\eta$ and the Fock space $H_{\psi_0}$ of the free fermion? Or simply, does $K$ contain free fermion states? The answer is no. The proof of this statement can be found in [2, 21, 22], with minor adaptations for implementation in curved spacetime. Let $\mathcal{F}_0' \subset \mathcal{F}'$ be the sub-algebra that commutes with $J_\mu^L$. We write the solutions of the subsidiary condition in $K$ as $K' = \overline{(\mathcal{F}_0'(\Sigma, \psi')\Omega)}$, (where we have denoted $\mathcal{F}_0'(\Sigma, \psi')$ as the field algebra generated by $\Sigma$ and $\psi'$). The physical space is

$$H_{phys} = \overline{K'/K''} \equiv \bar{H}_\Sigma \otimes \bar{H}_{\eta,\psi_0} \tag{41}$$

where $K''$ is the null subspace of $K'$. We now have the following

**Theorem:** The electric charge vanishes on the physical space.

*Sketch of Proof:* (adapted to curved spacetime from Strocchi [2], methodology from [23])

Let $Q_\rho^{el} = \int j^0(x) f_\rho(x_1)\alpha(x_0)\sqrt{|g|}dx_0 dx_1$. We prove the theorem by showing that for any vector $\hat{\Psi} \in H_{phys}$; $s-\lim_{\rho\to\infty} \exp[i\lambda \hat{Q}_\rho^{el}]\hat{\Psi} = \hat{\Psi}$.

Proof: Let $\Omega$ be the vacuum for the physical space (e.g. the unit element). Use the carat symbol to represent operators on the physical space. The operator solution of the Schwinger model is given by equation (11) with the ansatz $\varphi \to i\pi\gamma^5(\eta+\Sigma)/e$. The Maxwell equation

$$F^{\mu\nu}{}_{;\nu} = -eJ^\mu + a^\mu \quad \text{with} \quad a^\mu = eJ^\mu_{free} - \frac{e}{\sqrt{\pi|g|}}\varepsilon^{\mu\nu}\partial_\nu\tilde{\eta}$$

together with the subsidiary condition $\nabla_1 \hat{F}^{01} = -e\hat{J}^0$ leads to the identity $\hat{Q}^{el}_\rho = \frac{-1}{\sqrt{-g}}\partial_1[\sqrt{-g}\hat{F}^{01}(f_\rho\alpha)]$; and note that we also have

$$\frac{1}{\sqrt{|g|}}\partial_1[\sqrt{|g|}F^{01}(f_\rho\alpha)] = \frac{1}{\sqrt{|g|}}\frac{e}{\sqrt{\pi}}\partial_1[\Sigma(f_\rho\alpha)] \equiv Q^\Sigma_\rho.$$

Here $f$ and $\alpha$ are in accordance with equations (15) and (16).

<u>Step 1.</u>     We will further assume that $\lim_{\rho\to\infty}\|Q^\Sigma_\rho\Omega\| = 0$.

In the limit this merely expresses the property that the vacuum is a simultaneous eigenstate for all continuous symmetry groups. *However*, this property is only necessarily true for free fields, in Minkowski space. The $\Sigma$ field is massive as a result of interacting with a charged fermion field; in Minkowski space the above assumption may be proved. The proof depends on global properties of spacetime. In our case, the proof has proven extremely difficult. Nevertheless, in the sequel it is shown that $\Sigma$ *is* a free field, albeit a massive one; the property is true and provable in Minkowski space; and one hopes that in some limit as the curvature vanishes Minkowski physics resumes.

<u>Step 2.</u>     Show $[1 - \text{s-}\lim_{\rho\to\infty}\exp(i\lambda\hat{Q}^{el}_\rho)]D_\Sigma\Omega = 0$ where $D_\Sigma$ is the set of vectors resulting from the application of polynomials of $\tilde{\Sigma}$ to $\Omega$. $D_\Sigma$ is dense in $H_{phys}$ and it is a set of analytic vectors for $\frac{1}{\sqrt{|g|}}\frac{e}{\sqrt{\pi}}\partial_1[\Sigma(f_\rho\alpha)] \equiv Q^\Sigma_\rho$. Let $\Psi$ belong to a finite particle subspace of $\bar{H}_\Sigma$.

Standard estimates of $\left\|\left(\frac{1}{\sqrt{|g|}}\frac{e}{\sqrt{\pi}}\partial_1[\Sigma(f_\rho\alpha)]\right)^n\Psi\right\|$ insure that $\text{s-}\lim_{\rho\to\infty}(i\lambda\hat{Q}^{el}_\rho)D_\Sigma\Omega = 0$.

Therefore $\|\Omega - \exp(i\lambda Q^\Sigma_\rho)\Omega\| \to 0$ as well. Step 1 has asserted that $\lim_{\rho\to\infty}\|Q^\Sigma_\rho\Omega\| = 0$ and by the estimate $|e^{ix} - 1| \leq |x|$ we have that $[1 - \exp(i\lambda Q^\Sigma_\rho)]D_\Sigma\Omega = 0$ in the same limit. Since $[1 - \exp(i\lambda Q^\Sigma_\rho)] = A_\rho$ is a bounded operator, the strong limit of $A_\rho$ vanishes on the whole Fock space of $\tilde{\Sigma}$, and therefore on the physical space, so that $[1 - \text{s-}\lim_{\rho\to\infty}\exp(i\lambda\hat{Q}^{el}_\rho)]D_\Sigma\Omega = 0$ as was to be proven. The Theorem follows by the characterization of the physical space as shown in equation (41), since $\tilde{\Sigma}$ and $\mathcal{F}_0'$ commute. ∎

The structure we have introduced is somewhat richer than it may at first appear. The Schwinger model in Minkowski spacetime has long been a laboratory for displaying the richness of the vacuum structure of gauge theories. This work has carried forth into curved spacetime investigations [4]. Here we will only make the following observations:

We have labeled as $\mathcal{F}$ the algebra generated by the charged field $\psi'$ and by $A_\mu$. Here the spinor $\psi'$ is defined by $\psi'(x) =: e^{i\sqrt{\pi}\gamma^5\eta} : (x)\psi_{free}(x)$ or $\psi'(x) =: e^{i\sqrt{\pi}\gamma^5\eta} : (x)e^{-\frac{1}{2}\sigma}\psi_0(x)$; where the last expression is in the "isothermal" coordinates we have been using. The "vacuum" $\Omega$ obtained by the GNS procedure from the physical space provides an irreducible representation for the correlation functions obtained from $\mathcal{F}$ but a *reducible* representation of the sub-algebra of local observables (call it $\mathcal{A}_{obs}$). $\mathcal{A}_{obs}$ is generated by $F^{\mu\nu}$ and the bi-local operators [6] $T(x,y)$ which we will now define. For $\psi$ obtained from equation (45) one defines the gauge invariant operators $T(x,y) = \psi(x)\exp[ie\int_s A^\mu(\xi)d\xi_\mu]\psi^\dagger(y)$, where $s$ is along the geodesic connecting $x$ and $y$. On the physical space we use $\hat{T}(x,y) = \hat{\psi}(x)\exp[ie\int_s A^\mu(\xi)d\xi_\mu]\hat{\psi}^\dagger(y)$. We will show that the algebra $\hat{\mathcal{A}}_{obs}$ operating on the physical space has a *non-trivial center*.

To begin, the general form of the massless, scalar two-point function is usually defined $\langle\varphi(x)\varphi(y)\rangle = -i[V(x,y)|\ln\bar{\sigma}+i\varepsilon| + W(x,y)]$ [see appendix B] where $2\bar{\sigma}$ is the square of the geodesic distance between $x$ and $y$, and $V$ and $W$ are defined as in reference [13]. With respect to the definition of normal products as in equation (40), one may amend this and define the Hadamard paramatrix as defined by Wald in [9]:

$$H_W(x,y) = V(x,y)\left|\ln\bar{\sigma} + 2i\varepsilon(x_0 - y_0) + \varepsilon^2\right| + W(x,y) \tag{42}$$

To render the argument of the logarithm dimensionless one can re-write $\langle\phi\phi\rangle$ as

$\langle\phi\phi\rangle(x) = \lim_{x\to y}\{V(x,y)\ln\left|M^2\bar{\sigma} + 2i\varepsilon(x_0 - y_0) + \varepsilon^2\right| + W'(x,y)\}$ where $W'(x,y) = W(x,y) - V\ln M^2$ for the massless scalar and for an arbitrary renormalization constant $\frac{1}{2}M^2$. To define the "time ordered product" as in equation (31) and following reference [16], one drops the $W$ term in the two point function $\langle\phi\phi\rangle(x) = \lim_{x\to y}\{V(x,y)\ln\left|2M^2\bar{\sigma} + i\varepsilon\right| + W'(x,y)\}$ and expresses the covariant singular part as the associated Hadamard parametrix

$$H(x,y) = V(x,y)|\ln \bar{\sigma} + i\varepsilon| \tag{43a}$$

The $W(x,y)$ term is state dependent, and can be picked up later when defining two point functions and Green functions. The expression (43) is covariant and independent of the state definition. Conditions on OPE coefficients will suffice to render the $2i\varepsilon$ term unnecessary. For the two dimensional version of equation (42) let us consider $\lim_{\bar{\sigma}\to 0} V(x,y)$. Now in this limit we have from

[13] that $V \to \dfrac{1}{4\pi}\left[-1 + \dfrac{1}{2}(-m^2 + \xi_c + \dfrac{1}{6}R)\bar{\sigma}\right]$. In 2D, $\xi_c = 0$. \quad (43b)

In the following argument, we will use bosonization to describe massless spinor fields. The reason is that two point functions are quite easy to manipulate and the cancellations that occur become obvious. Consider a solution to the free Dirac equation $e^{-\sigma/2}\begin{pmatrix}\psi_{10}(x)\\\psi_{20}(x)\end{pmatrix}$. Let us define (from [21]) the components as follows:

$$\psi_{10} = \dfrac{1}{\sqrt{2\pi}} e^{-i\frac{\sqrt{\pi}}{4}Q_R} :e^{2i\sqrt{\pi}\varphi_L}:(x) \text{ and } \psi_{20} = \dfrac{1}{\sqrt{2\pi}} e^{i\frac{\sqrt{\pi}}{4}Q_L} :e^{2i\sqrt{\pi}\varphi_R}:(x) ; \tag{44}$$

and here $\varphi$ is any massless scalar field. In the Schwinger model, we use $\varphi = \eta$,

Here, the possibility of charge in the physical space is introduced via the field $\eta$, which we are allowed to do by gauge invariance, and thus exploit gauge invariance to facilitate the bosonization. The dynamics of the theory finally render the charge absent in the physical space.

Reference [21] further asserts that $Q_{R,L} \equiv i\pi(\varphi_{+,R,L}(v_0) - \varphi_{-,R,L}(v_0))$ ; and since $\varphi_{R,L} = \dfrac{1}{2}(\varphi \pm \tilde{\varphi})$ we have $[\varphi_{R,L}, Q_{R,L}] = -i$. In curved spacetime the dual is defined $\partial_\mu \tilde{\varphi} = \sqrt{-g}\partial^\nu \varepsilon_{\mu\nu}\varphi$. With these definitions, equation (32) and the formula
$\{\Psi_0 : \exp(g_1\phi):(x_1):\exp(g_2\phi):(x_2)\Psi_0\} = \exp(g_1 g_2 H(x_1,x_2))$ adapted from [7], let us consider the object $\mathcal{T}(x) = e^{-\sigma}:[\psi_{01}^\dagger : e^{i\sqrt{\pi}\eta}:(x)][:e^{i\sqrt{\pi}\eta}:(x)\psi_{02}]:(x)$. Since right and left operators commute, this object commutes with itself at different points. Thus $\hat{\mathcal{A}}_{obs}$ has a non-trivial center. Now define $\zeta(x) \equiv e^{\sigma(x)} 2\pi \hat{\mathcal{T}}(x)$; where the carat implies restriction of the operator to the physical space. With chiral invariance of the vacuum
$\langle \mathcal{T}(x_1)...\mathcal{T}(x_n)\mathcal{T}^\dagger(x_1)...\mathcal{T}^\dagger(x_m)\rangle = 0$ for $n \neq m$. For $n = m$, one notices that the correlation functions of $:\psi_{01}^\dagger \psi_{02}:$ are cancelled by those of $:e^{i2\sqrt{\pi}\eta}:(x)$; a result of the choice of negative sign for the two point function of $\eta$. Now in the limit as $\bar{\sigma} \to 0$ we have $\bar{\sigma} \approx (x-y)^2/2$ and $\zeta^*(x)\zeta(y) = \zeta(x)\zeta^*(y) = 1$; as well as $\langle \zeta(x_1)...\zeta(x_n)\zeta^*(y_1)\zeta^*(y_m)\rangle = \delta_{n,m}$.

Consider the chiral transformations $\psi \to e^{-i\alpha\gamma^5}\psi \qquad A_\mu \to A_\mu$. On the physical space these transformations induce $\zeta \to e^{-2i\alpha}\zeta \qquad \hat{\tilde{\Sigma}} \to \hat{\tilde{\Sigma}}$.

NB The eigenvalues of the $\gamma^5$ operator are easily obtained from $\langle \bar\psi \tfrac{1}{2}(1+\gamma^5)\psi\rangle = 0$, which implies that $\gamma^5 \psi_{eig} = 2\psi_{eig}$, and therefore the chirality changes by a factor of 2 under these transformations. We can trivialize the center for $\mathcal{A}_{obs}$ by introducing

$$|\theta\rangle = \sum e^{-2in\theta}\,|2n\rangle. \tag{45}$$

Here $|2n\rangle \equiv \hat{\Psi}_{0n} = (\zeta)^{2n}|\Omega\rangle$ and $Q^5 \hat{\Psi}_{0n} = 2n\hat{\Psi}_{0n}$ for $Q^5 = \lim_{r\to\infty}\int \bar\psi\gamma^0\gamma^5\psi(f,\alpha)\sqrt{-g}\,d^2x$. Using this nomenclature the operators $\zeta = e^{-2i\theta}$ become C-numbers on $|\theta\rangle$.

## INTRODUCTION TO THE OPERATOR PRODUCT EXPANSION

We have introduced the operator solution of the Schwinger model, defined the Hilbert space generated by a GNS procedure with its associated "vacuum", suitably enlarged the physical space to allow for the possibility of charged states and introduced a modified vacuum to handle the non-trivial center generated by local chiral-density operators. The OPE formalism introduced by Wald and Hollands is specifically designed to handle a complete solution generated by a perturbation expansion, and the purpose of our exercise here is to show how this might be translated to a theory that is exactly solvable.

To begin: introduce three free quantum fields $\Sigma, \eta$ and $\psi$. These are taken to exist in a two dimensional, globally hyperbolic manifold $M$ and have the character $\Sigma$ = massive scalar, $\eta$ = a massless scalar field quantized so that its two point function has the opposite sign normally associated to a massless scalar field, and finally $\psi$ = a massless "spinor" field; which is to say a two component object that obeys a field equation $i\nabla\!\!\!\!/\,\psi = 0$ in $M$, with notation as above. Now $\psi$ and its conjugate $\bar\psi$ are quasi-free and defined as above following equation (3C). The fields $\Sigma$ and $\eta$ are also quasi-free and defined as follows equation (2C), with conditions (a) to (d'). Using the above conventions, let us also define the *algebras $\mathcal{A}(\tilde\Sigma)$, $\mathcal{A}(\tilde\eta)$ and $\mathcal{A}(\psi,\bar\psi)$. Notation indicates that $\eta \to \tilde\eta$ has been enlarged by the Krein procedure, and $\tilde\Sigma$ is the dual field for $\Sigma$. Using standard quantization rules for fields in curved spacetime, one may define states and apply the GNS theorem to provide a Hilbert space, and hence a Fock space for each of the fields.

Since all 2D spaces are conformally flat, we can generally utilize conformal coordinates and write the metric as in equation (8) $ds^2 = e^{2\sigma}(dx_0^2 - dx_1^2)$. In 2D spacetime one can always define a massless field as $\tilde{\partial}_\mu \eta \equiv \partial_\mu \tilde{\eta}$. We will write the field $A_\mu(x) = -\frac{\sqrt{\pi}}{e}(\tilde{\partial}_\mu \Sigma(x) + \tilde{\partial}_\mu \eta(x))$ and the field $F_{\mu\nu} = -\frac{\sqrt{\pi}}{e}(\partial_\mu \tilde{\partial}_\nu - \partial_\nu \tilde{\partial}_\mu)\Sigma = -\frac{\sqrt{\pi}}{e}\varepsilon_{\mu\nu}\sqrt{-g}\Delta\Sigma$. We have thus defined a vector field $A_\mu(x)$ and a skew symmetric tensor $F_{\mu\nu}$ in terms of our basic fields. In addition, we also require $\Delta(\Delta + \frac{e^2}{\pi})\Sigma = 0$. These relations will form an equivalence, along with a definition of the physical Hilbert space, between the fundamental fields $\Sigma, \eta$ and $\psi$ and the usual fields of electrodynamics. From the last relation we may write $F_{\mu\nu} = \frac{e}{\sqrt{\pi}}\eta_{\mu\nu}\Sigma$. We can thus replace quantization for $F_{\mu\nu}, A_\mu$ with $\eta_{\mu\nu}\Sigma, \tilde{\partial}_\mu \Sigma$, on the *physical space*. In the prequel we have seen that $\tilde{\partial}_\mu \eta$ and $J_\mu^{free}$ are not in the physical space. The fundamental Fermion fields in this context will be $\psi'(x) = e^{-\sigma/2} : e^{i\sqrt{\pi}\gamma^5 \eta} : (x)\psi_0(x)$ and $\bar{\psi}'(x) = e^{-\sigma/2}\bar{\psi}_0(x) : e^{-i\sqrt{\pi}\gamma^5 \eta} : (x)$. These will be the basic fields with which the theory can be defined. The final interacting fields on the physical space are then: $F_{\mu\nu} = \frac{e}{\sqrt{\pi}}\eta_{\mu\nu}\Sigma$, $A_\mu(x) = -\frac{\sqrt{\pi}}{e}\tilde{\partial}_\mu \Sigma(x)$ and $\psi x) =: e^{i\sqrt{\pi}\gamma^5 \Sigma} : (x)\psi'(x)$ and $\bar{\psi}(x) = \bar{\psi}'(x) : e^{-i\sqrt{\pi}\gamma^5 \Sigma} : (x)$. These are the objects to appear in Operator Product Expansions.

The first point to address is the form of the "normalized" two point function we will select for a scalar field of mass *m*. From [20, 24], one gets the general expression for the two dimensional two point function for a massive field

$$G^+(x,x') = \frac{\Delta^{1/2}(x,x')}{4}\sum_{j=0}^{\infty} a_j(x,x')\left(\frac{-\partial}{\partial m^2}\right)^j H_0^{(2)}(\sqrt{2m^2\sigma}) \quad , \tag{46}$$

the same structure introduced in the no-charge theorem. This expression is also written

$$G_{DS}^F(x,x') = \frac{1}{4}\sum_{n=0}^{\infty} \frac{(-)^n}{(m^2)^n} A_n(x,x')\left(\frac{z(x,x')}{2}\right)^n H_{-n}^{(2)}(z(x,x')) \tag{47}$$

[see Appendix B]. Appendix B shows how this may be put into the Hadamard form

$H(x,x') = V(x,x')\ln|\bar{\sigma} + i\varepsilon| + W(x,x')$ and defines $V$ and $W$. With our conventions, then, the two point function is $G^+(x,x') = -i[V(x,y)\ln|\bar{\sigma} + i\varepsilon| + W(x,y)]$ for $x_0 > x'_0$.

The algebra $\mathcal{A}(M)$, may be considered the product of the *-algebras based on $\mathcal{A}(\tilde{\Sigma}), \mathcal{A}(\tilde{\eta})$ and $\mathcal{A}(\psi,\bar{\psi})$. Here $\psi$ is the free, massless Dirac field $e^{-\sigma/2}\psi_{free}$. Appendix C shows how to define the state spaces $\mathcal{S}(\tilde{\Sigma}), \mathcal{S}(\tilde{\eta}), \mathcal{S}(\bar{\psi},\psi)$ of these algebras generated by monomials of the fields.

We may thus add the fields $\int :e^{i\sqrt{\pi}\gamma^5\eta}:(x)f(x)\sqrt{-g}\,dx$ and $\int :e^{i\sqrt{\pi}\gamma^5\Sigma}:(x)f(x)\sqrt{-g}\,dx$ to $\mathcal{S}(\tilde{\Sigma})$ and $\mathcal{S}(\tilde{\eta})$. With these, we can add the smeared product $\psi(x) = \int :e^{i\sqrt{\pi}\gamma^5\eta}:(x)\psi(x)f(x)\sqrt{-g}\,dx$ to $\mathcal{A}(\psi,\bar{\psi})$. Let $\mathcal{S}'(\psi,\bar{\psi}) = \mathcal{S}(\psi,\bar{\psi}) - \Omega_\psi$ where $\Omega_\psi$ is the "vacuum" for $\mathcal{A}(\psi,\bar{\psi})$. Let $\mathcal{S}'(\tilde{\eta}) = \mathcal{S}(\tilde{\eta}) - \Omega_{\tilde{\eta}} - \mathcal{M}(\tilde{\eta})$ where $\mathcal{M}(\tilde{\eta})$ is the set of all states formed by powers of $\tilde{\eta}$ and its derivatives $\partial^n_{\mu_1...\mu_n}\tilde{\eta}$ on $\Omega_{\tilde{\eta}}$. Then the total state space will be

$$\mathcal{S}(M) = \mathcal{S}(\tilde{\Sigma}) \otimes \mathcal{S}(\tilde{\eta}) \otimes \mathcal{S}(\psi,\bar{\psi})/(\mathcal{S}'(\psi,\bar{\psi}) \cup \mathcal{S}'(\tilde{\eta})). \tag{48}$$

Equation (48) means that, except for the vacuum, there are no states associated with operator terms like $X\psi Y$ or $X\bar{\psi}Y$ and there are no individual finite power terms in $\tilde{\eta}$.

In the following we always make the identifications:

$$A_\mu = -\tilde{\partial}_\mu \Sigma - \tilde{\partial}_\mu \eta; \quad F_{\mu\nu} = -\frac{\sqrt{\pi}}{e}\eta_{\mu\nu}\Delta\Sigma = \frac{e}{\sqrt{\pi}}\eta_{\mu\nu}\Sigma. \tag{49}$$

This allows an unambiguous definition of the reduced algebra of operators in terms of the auxiliary fields.

**OBSERVATION:** *The stress energy tensor for the interacting field represents that of a free, massive vector field.*

We begin by calculating the stress-energy tensor on the physical space. This is composed of two terms, $T^{\mu\nu} = -\hat{F}^{\mu\lambda}\hat{F}^\nu_\lambda + \frac{g^{\mu\nu}}{4}\hat{F}^2 + i\hat{\bar{\psi}}\gamma^\mu\nabla^\nu\hat{\psi}$. The third term is symmetrized as follows:

$$i\hat{\bar{\psi}}\gamma^\mu\nabla^\nu\hat{\psi} \Rightarrow \frac{i}{4}\{\hat{\psi}^\dagger\hat{\gamma}^0\gamma^\mu(D^\nu - ieA^\nu)\hat{\psi} - [(D^\nu + ieA^\nu)\hat{\psi}^\dagger]\hat{\gamma}^0\gamma^\mu\hat{\psi}\} + (\mu \leftrightarrow \nu) \text{ with}$$

$D^\nu = \partial^\nu - \frac{1}{2}e\omega^\nu_{ab}\Sigma^{ab}$. The first two terms reduce to $-\frac{g^{\mu\nu}}{4}\hat{F}^2$ in 2 dimensions. Because of (49) we

have $\qquad :F^2(x): = -2m^2:\Sigma^2(x): \tag{50}$

and using the definition of $\hat{T}(x, y) = \hat{\psi}(x) \exp[ie \int_s A^\mu(\xi) d\xi_\mu] \hat{\psi}^\dagger(y)$ we will write

$$\hat{\psi}^\dagger \hat{\gamma}^0 \gamma^\mu (D^\nu - ieA^\nu) \hat{\psi} = -\lim_{\substack{y \to x \text{ on } s \\ (x-y)^2 \neq 0}} D_x^\nu \hat{T}(x, y) \tag{51}$$

Equation (50) is defined by the expression (37). With the preceding, the stress energy becomes

$$T^{\mu\nu} = \frac{1}{2} m^2 g^{\mu\nu} : \Sigma^2(x) : -\frac{i}{4} \lim_{\substack{y \to x \text{ on } s \\ (x-y)^2 \neq 0}} \left[ (\partial_x^\nu - \partial_y^\nu) \left\{ \hat{\gamma}^0 \gamma^\nu [\hat{T}(x, y) - \underset{x \to y}{\text{sing}}(\hat{T}(x, y))] \right\} + (\mu \leftrightarrow \nu) \right]. \tag{52}$$

This expression requires some explanation. The symmetrization procedure has reduced the operator $D^\nu$ to $\partial^\nu$. From $\tilde{\omega}(x, y)$ as defined in the discussion preceding equation (34), one extracts the two-point function $\omega^+(x, y) = i\nabla \tilde{\omega}$. This will have the singular structure given by $i\nabla [V(x, y) \ln(\bar{\sigma} + i\varepsilon)]$, which is the singularity structure of $\hat{T}(x, y)$. This term, rather than $\langle \Omega | \hat{T}(x, y) | \Omega \rangle$, is the term subtracted from $\hat{T}(x, y)$. From this point, the calculation proceeds exactly as in [28], with the result (absent the trace anomaly)

$$T^{\mu\nu} =: \left\{ \partial^\mu \Sigma \partial^\nu \Sigma - \frac{1}{2} g^{\mu\nu} \left[ (\partial_\rho \Sigma)(\partial^\rho \Sigma) - m^2 \Sigma^2 \right] \right\}(x) : \tag{53}$$

i.e. a free, massive scalar field.

On the physical space we always use $\frac{e}{\sqrt{\pi}} \hat{A}_\mu = -\tilde{\partial}_\mu \Sigma$. The field $F_{\mu\nu}$ can then be defined with derivatives. Let us consider the simplest case, the product of two vector fields $\hat{A}_\mu(x) \hat{A}_\nu(y)$. This will serve to motivate an additional consideration that is peculiar to the two dimensional case. Anticipating Wick's theorem one writes

$$\frac{e^2}{\pi} \hat{A}_\mu(x_1) \hat{A}_\nu(x_2) = \tilde{\partial}_\mu \Sigma(x_1) \tilde{\partial}_\nu \Sigma(x_2) =: \tilde{\partial}_\mu \Sigma(x_1) \tilde{\partial}_\nu \Sigma(x_2) :_H + \Gamma_{\mu\nu}(x_1, x_2) \tag{54}$$

We have defined $\Gamma_{\mu\nu}(x_1, x_2) \equiv \left( \frac{e^2}{\pi} \hat{\delta}_{\mu\nu} + \tilde{\nabla}_\mu \tilde{\nabla}_\nu \right) \left[ V(x_1, x_2) \ln \left| M^2 \bar{\sigma} + i\varepsilon \right| \right]. \tag{55}$

Here the notation is $\hat{\delta}_{\mu\nu} = \begin{cases} 1 & \mu = \nu = 0 \\ -1 & \mu = \nu = 1 \\ 0 & \mu \neq \nu \end{cases}$. The function $V$ is the same as that which appears in the scalar field case. We have included the factor $M^2$ introduced in equation (43). An issue mentioned in [1] is connected with the asymptotic positivity requirement for fields in two dimensions. We are only concerned with one field, $\hat{A}_\mu = -\tilde{\partial}_\mu \Sigma$ on the physical space. But we deal with the derivative always, and this eliminates the logarithmic singularity that is problematic in their proofs.

Using Riemann normal coordinates for $x_1$ and $x_2$, centered at $y$, $x_1, x_2$ can be expressed as a Taylor series around $y$. However, one requires the geodesic distance running through $x_1, x_2$ but referred to the diagonal at $y$. This is given by $L_{12}^2 = g_{\mu\nu}\Delta_{12}^\mu \Delta_{12}^\nu - \tfrac{1}{3} R_{\mu\alpha\nu\beta} x_1^\alpha x_1^\beta \Delta_{12}^\mu \Delta_{12}^\nu + \mathcal{O}(\varepsilon^3)$ where $\Delta_{12}^\mu = x_2^\mu - x_1^\mu$. Here, $\varepsilon$ is of the size of the open set inside which the Riemann coordinates are defined and $y$ is the diagonal. The coefficient of the identity defines the strength of the singularity when $s \to 0$.

On the physical space, equation (33) becomes

$$F^{\mu\nu}{}_{;\nu} = \frac{1}{\sqrt{-g}} \frac{\partial}{\partial x^\nu}(\sqrt{-g} F^{\mu\nu}) = -m^2 A^\mu \quad \text{with} \quad m^2 = \frac{e^2}{\pi} \tag{56}$$

These are curved spacetime Proca equations for a massive vector field, and one may consult the quantization procedure in [29].

With the formal example of equation (54), let us define the precise notion of an operator product expansion as given in [1]. On the physical space, other than the vacuum, the only objects that occur are elements of the Proca field, and it is a free field. OPEs will therefore only need to be formed from the fields $F_{\mu\nu}$ and $A_\mu$, since all others can be formed by combinations of these by using equation (56) and the relations involving the $c^{(i)}$. These objects are quasi-free, and this transforms multiple powers into simple expressions. For the three-fold product one has

$$\frac{e^3}{\pi^{3/2}} \hat{A}_\mu(x_1)\hat{A}_\nu(x_2)\hat{A}_\rho(x_3) = \frac{e^3}{\pi^{3/2}} :\hat{A}_\mu(x_1)\hat{A}_\nu(x_2)\hat{A}_\rho(x_3):_H + \frac{e}{\pi^{1/2}} \hat{A}_\mu(x_1)\Gamma_{\nu\rho}(x_2, x_3) + cyclic$$

$$= \frac{e^3}{\pi^{3/2}} \hat{A}_\mu(x_1)\hat{A}_\nu(x_2)\hat{A}_\rho(x_3) =: \tilde{\partial}_\mu \Sigma(x_1)\tilde{\partial}_\nu \Sigma(x_2)\tilde{\partial}_\rho \Sigma(x_3):_H + \tilde{\partial}\Sigma_\mu(x_1)\Gamma_{\nu\rho}(x_2, x_3) + cyclic \tag{57}$$

or Wick was wrong a long time ago. Before proceeding, we will borrow two formulae from [30]. The local covariant n-th Wick power as a distribution valued over a field algebra $\{\phi\}$ is given by

$$\phi^n(x) = \lim_{\varepsilon \to 0} : A^{\otimes n} :_H (x'(\varepsilon)) \quad x'(\varepsilon) = (\exp_x(\varepsilon \xi_1),...,\exp_x(\varepsilon \xi_n)) \tag{58}$$

Where $\xi_i$ denote the Riemann normal coordinates of the point $x_i$ relative to $x$. [NB, as Hollands shows, this definition is covariant]. Also, on a manifold $M^n$

$$: \phi^{\otimes n} :_H (x_1,...,x_n) :: \phi^{\otimes m} :_H (x_{n+1},...,x_{n+m}) =$$

$$\sum_k \frac{n!m!}{(n-k)!(m-k)!k!} \sum_{p_1,...p_k \in P} \prod_i \Gamma(x_{p_i(1)} x_{p_i(2)}) : \phi^{\otimes n+m-2k} :_\omega (\{x_j; j \notin |P|\}) \tag{59}$$

is a statement of Wick's theorem for normal ordered quantities relative to a state with two point function $H = \Gamma$ as defined in equation (31). $P$ is the set of all pairs $p_i \in \{1,...,n\} \times \{n+1,...n+m\}$.

In the formalism of [1, 30] it is much easier to produce the OPE coefficients than to explain why the objects produced are the correct ones. In Wilson's original formulation, operator products at different nearby points are expressed in terms of operators at a reference point and the product terms are obtained by a Taylor expansion similar to equation (55). The singularity at the reference point is a result of the physics of the problem and a key source of information regarding the product term. The methodology is similar in curved space time, but there are obvious issues of "covariant distances" to deal with and the formalism is constructed to take advantage of covariant terms, e.g. the variable $\sigma(x_1, x_2)$, and a sophisticated construction of "nearness" to insure control of remainder terms in the series in a unique manner. Fortunately, for free fields the apparatus is quite simple, as all issues of defining an interacting theory perturbatively are avoided.

From [1], the nomenclature for the operator product expansion is given as

$$\left\langle \phi^{(i_1)}(x_1)...\phi^{(i_n)}(x_n) \right\rangle_\omega \approx \sum_j C^{(i_1)...(i_n)}_{(j)}(x_1,...,x_n; y) \left\langle \phi^{(j)}(y) \right\rangle_\omega \tag{60}$$

The points $x_i$ are imagined as vectors expressed in Riemann normal coordinates about a point $y$. Thus, from (57) one may read off the result from the RHS.

$C(x_1, x_2, x_3; y) = \frac{e}{\sqrt{\pi}} \Gamma_{\mu_i \mu_k}(x_i, x_k)$ $(i \neq k \neq j) = 1,2,3$. These simple examples show the relation of familiar expressions to the expressions in (58) and (59). In the free field case one may use equation (59) as a general expression for generating any operator expansion coefficient. But first one must examine the equivalence relation in equation (60).

Hollands' and Wald's formalism in [1] is designed to define an interacting field in curved spacetime by its perturbation series. Thus the equivalence is regarded as some type of strong convergence in an asymptotic limit as $x_1,...,x_n \to y$. Besides truncating the series at some specified level of accuracy in terms of products of fields, there are also issues of remainder terms

(as in a Taylor series) whose absolute value vanishes in some limit. In our case these terms involve curvature. And the "distances" between multiple points becomes rather involved in a general spacetime. Consider an expression for $C_j^{i,k}(x_1, x_2,..., x_4; y)$. We assume the reader is familiar with the constructions given in [1] and [30]. We've borrowed an illustration from reference [30] for clarity here.

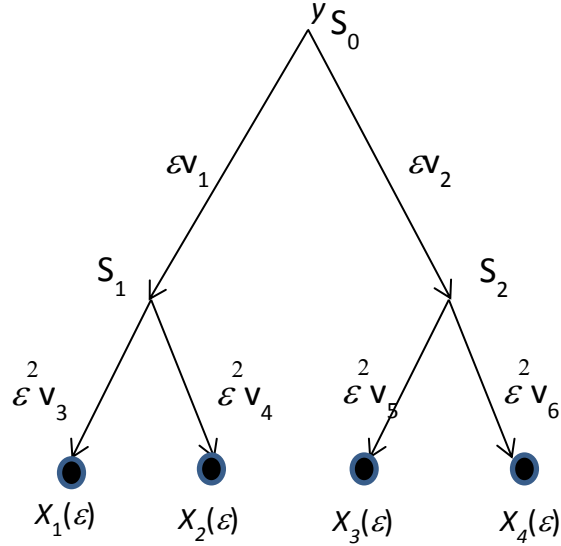

A particular tree $T = \{S_0, S_2,..., S_6\}$ for 4 points is shown, and a generalization for $n$ points seems obvious. Each point $x(\varepsilon)$ is given by a curve from the root $S_0$ by a unique path. With all the definitions from the references, we write $x_1(\varepsilon) = \varepsilon v_1 + \varepsilon^2 v_3$, $x_2(\varepsilon) = \varepsilon v_1 + \varepsilon^2 v_4$, $x_3(\varepsilon) = \varepsilon v_2 + \varepsilon^2 v_5$, $x_4(\varepsilon) = \varepsilon v_2 + \varepsilon^2 v_6$ and so on. There are many possible paths from $S_0$ to the $x_i$s, only one of which is shown above. Each possibility gives a different tree. If the vectors $v_i \neq v_j$, then all the points do not lie on the diagonal $y$. By this mechanism the points may be made to approach $y$ at different rates. This apparatus was designed by Fulton [32], but a very clear presentation for these purposes is given in [31]. From this one can imagine a construction for any $C_j^{i,k}(x_1, x_2,..., x_n; y)$. In [31] Holland shows that for free Boson fields, the general associativity condition $C_{(j)}^{(i)}(x_1, x_3,..., x_n; y) = \sum_{(k)} C_{(k)}^{(i)}(x_1, x_2,..., x_m; y) C_{(j)}^{(k)}(x_m, x_{m+1}, x_{m+2},..., x_n; y)$ may be proved by the symmetry condition $C_j(x_1, x_2; y) = C_{(j)}(x_2, x_1; y)$ along with the three point associativity relation $C_j(x_1, x_2, x_3; y) = \sum_k C_{(j)}(x_1, x_k; y) C_{(j)}(x_k, x_3; y)$ for the simplest coefficients, as in equations (55) and (57).

Now we will write a general expression for the operator product expansion in terms of the vector fields $A_\mu$. The notation is that introduced in appendix F.

The basic entities are free Boson fields with commutation relations $[A_\mu(x), A_\nu(y)] = 0$ for spacelike separations. The products $C_{(k)}(x, z; y)$ are therefore symmetric. The same relations insure that the simple associativity relation $C_j(x_1, x_3; y) = C_{(j)}(x_1, x_3; y) C_{(j)}(x_1, x_3; y)$ is satisfied. Therefore we may take it as true that the $C$'s are generally associative as well. Since the derivative is a linear operator, it will cause no difficulty for the associativity condition. Let us write a general OPE for two points:

$O_{\{\mu\}}\{A^i_\alpha(x_1)\} O_{\{\nu\}}\{A^j_\beta(x_2)\} \approx \sum_{\substack{k=\text{contractions} \\ \text{over } i,j}} C^\rho_{\alpha\beta}(x_1, x_2; y) O_{\{\mu\},\{\nu\}}\{A_\rho(y)\}$; where one understands the

contractions are such as to render all $k$ powers to linear terms. We have indicated some kind of an equivalence relation. If this were a strict equality, the first term would be $:O_{\{\mu\}}\{A^i_\alpha(x)\} O_{\{\nu\}}\{A^j_\beta(z)\}:$ and contractions would begin therefrom. Again, the first term would be $:O_{\{\mu\}}\{A^i_\alpha(y)\} O_{\{\nu\}}\{A^j_\beta(y)\}:$ with no coefficient. Taking the vacuum expection value of both sides would remove this term from consideration, and therefore one often sees the OPE relations written as $\langle O_{\{\mu\}}\{A^i_\alpha(x)\} O_{\{\nu\}}\{A^j_\beta(z)\}\rangle \approx \sum_{\substack{k=\text{contractions} \\ \text{over } i,j}} C^\rho_{\alpha\beta}(x, z; y) \langle O_{\{\mu\},\{\nu\}}\{A^k_\rho(y)\}\rangle$. Each contraction

introduces a Hadamard parametrix, as in

$:O_{\{\mu\}}\{A^i_\alpha(x)\}:_H :O_{\{\nu\}}\{A^j_\beta(z)\}:_H = :O_{\{\mu\}}\{A^i_\alpha(x)\} O_{\{\nu\}}\{A^j_\beta(z)\}:_H +$ (61)

$\sum_{1-\text{contraction } [(x,i)(z,j)]} \Gamma_{\mu_i \nu_j}(x, z) :O_{\{\mu\}}\{A^i_\alpha(x)\} O_{\{\nu\}}\{A^j_\beta(z)\}:_H +$

$\sum_{2-\text{contraction } [(x,i)(z,j)][(x,i')(z,j')]} \Gamma_{\mu_i \nu_j}(x, z) \Gamma_{\mu_i \nu_{j'}}(x, z) :O_{\{\mu\}}\{A^i_\alpha(x)\} O_{\{\nu\}}\{A^j_\beta(z)\}:_H + \ldots +.$

In our case this translates to $A^{n_1}_{\mu_{n_1}}(x_1) A^{n_2}_{\mu_{n_2}}(x_2) \cdots A^{n_k}_{\mu_{n_k}}(x_k) =: A^{n_1}_{\mu_{n_1}}(x_1) A^{n_2}_{\mu_{n_2}}(x_2) \cdots A^{n_k}_{\mu_{n_k}}(x_k):_H +$

$\sum_{1-\text{contraction } [(x_1, n_1)(x_2, n_2)\ldots(x_k, n_k)]} \Gamma_{\mu_{n_1} \nu_{n_2}}(x_1, x_2) :A^{n_1-1}_{\mu_{n_1-1}}(x_1) A^{n_2-1}_{\mu_{n_2-1}}(x_2) \cdots A^{n_k}_{\mu_{n_k}}(x_k): +$ (62)

$\sum_{2-\text{contraction } [(x_1, n_1)(x_2, n_2)\ldots(x_k, n_k)]} \Gamma_{\mu_{n_1} \nu_{n_2}}(x_1, x_2) \Gamma_{\mu_{n_2} \nu_{n_3}}(x_2, x_3) :A^{n_1-1}_{\mu_{n_1-1}}(x_1) A^{n_2-1}_{\mu_{n_2-1}}(x_2) A^{n_3-1}_{\mu_{n_3-1}}(x_3) \cdots A^{n_k}_{\mu_{n_k}}(x_k): + cyclic +$

The arcane notation is meant as (e.g. for $n_1 = 3$), $A^{n_1}_{\mu_{n_1}}(x_1) = A_{\mu_1}(x_1) A_{\mu_2}(x_1) A_{\mu_3}(x_1)$

This is a more cumbersome way of essentially repeating equation (59). With these two points of view one hopes that a general expression for an OPE will seem a bit more clear.

Because we are dealing with one variety of free field, we will begin with single powers of a quantity, and then show that the desired properties apply to the product of two powers, and use associativity to lead to the general case. First write

$$A_{\mu_1}(x_1)A_{\mu_2}(x_2)\cdots A_{\mu_k}(x_k) =: A_{\mu_1}(x_1)A_{\mu_2}(x_2)\cdots A_{\mu_k}(x_k):_H$$

$$+\sum_{perm}\Gamma_{\mu_1\mu_2}(x_1,x_2):A_{\mu_3}(x_3)\cdots A_{\mu_k}(x_k):+\sum_{perm}\Gamma_{\mu_1\mu_2}(x_1,x_2)\Gamma_{\mu_3\mu_4}(x_3,x_4):A_{\mu_5}(x_5)\cdots A_{\mu_k}(x_k):+$$

$$\mathbf{1}\cdot\sum_{perm}\Gamma_{\mu_1\mu_2}(x_1,x_2)\Gamma_{\mu_3\mu_4}(x_3,x_4)\cdots\Gamma_{\mu_{k-1}\mu_k}(x_{k-1},x_k)\text{ ; if }k\text{ is even.} \quad (63)$$

If $k$ is odd, the last term is written $\sum_{perm}\Gamma_{\mu_1\mu_2}(x_1,x_2)\Gamma_{\mu_3\mu_4}(x_3,x_4)\cdots\Gamma_{\mu_{k-2}\mu_{k-1}}(x_{k-2},x_{k-1})A_{\mu_k}(x_k).$ (64)

From this expressions, powers of the field may be considered by taking appropriate limits as the requisite variables approach each other; e.g. for a second power one has

$$A^2_{\mu_2}(x_1)=\lim_{v_3,v_4\to 0}A_{\mu_1}(y+\varepsilon v_1+\varepsilon^2 v_3)A_{\mu_2}(y+\varepsilon v_1+\varepsilon^2 v_4). \quad (65)$$

Obviously such limits depend upon how the tree is constructed.

Now the entire apparatus of references [1] and [30] is based on defining the interacting field as an asymptotic series in the polynomials of the basic, free fields. We have here an exact expression for the interacting field, using quasi-free fields as building blocks. Because of this, we have an exact expression for the two point functions, without having to specify an accuracy $\Delta$ by which we know when to terminate a perturbation series. In curved spacetime, this $\Delta$ also must take into account the curvature terms and the order to which they must be taken. Most of the structure introduced in the reference [30] can be discarded and we may simply use the Hadamard term (43a) *in toto*. From symmetry, and associativity for n = 3, we may thus write any power of an OPE in terms of expressions like equation (63). If one now takes the VEV of equation (63) the expansion coefficients result immediately as

$$\sum_{perm}\Gamma_{\mu_1}(x_1,x_2)\Gamma_{\mu_3}(x_3,x_4)\cdots\Gamma_{\mu_{k-2}}(x_{k-2},x_{k-1})\langle A_{\mu_k}(x_k)\rangle$$

$$\text{or }\mathbf{1}\cdot\sum_{perm}\Gamma_{\mu_1}(x_1,x_2)\Gamma_{\mu_3}(x_3,x_4)\cdots\Gamma_{\mu_{k-1}}(x_{k-1},x_k), \quad (66)$$

depending on whether the original product has an even or odd number of terms. A general expression such as $A^{n_1}_{\mu_1}(x_1)A^{n_2}_{\mu_2}(x_2)\cdots A^{n_k}_{\mu_k}(x_k)$ is reduced by pairs according to (63) until an expression such as (65) or (66) emerges. Thus one may use (66) as the model to examine the required properties OPEs must have in the Schwinger model.

On the physical space, the only field that is relevant to this discussion is $A_\mu(x)$, a massive, free field. It is nevertheless instructive to itemize the axioms and illustrate how the various constructions are meant to be applied in this simple case. The basic algebra would appear to be constructed from a massless vector field $A_\mu$ and a massless spinor field $\psi$. And yet the physical field is a massive vector and the spinors have disappeared in the finite particle sector of Hilbert space. The resultant free vector field obviously has a great deal more structure than a naïve first glance would presume, and this should be reflected in our choice of the original field algebra.

All of the apparatus necessary for a complete definition of the Schwinger Model by way of the Operator Product Expansion has now been introduced. Part II of this series will proceed with this matter and complete the argument.

Useful Formulae

$\hat{\Delta} = \partial_0^2 - \partial_1^2;\quad \Delta f = \dfrac{1}{\sqrt{-g}}\partial_i(\sqrt{-g}g^{ij}\partial_j f);\quad \gamma^5\gamma_\nu = \sqrt{-g}\gamma^\mu\varepsilon_{\mu\nu} = \gamma^\mu\eta_{\mu\nu};\quad \hat{\gamma}_0 = \hat{\gamma}^0\quad \hat{\gamma}_1 = -\hat{\gamma}^1;$

$\tilde{\nabla}_\mu \equiv \sqrt{-g}\varepsilon_{\mu\nu}\nabla^\nu;\quad \varepsilon^{\nu\sigma}\hat{\gamma}_\sigma = \hat{\gamma}^\nu;\quad \hat{\gamma}^0\gamma^5 = \gamma^1 = -\gamma^5\hat{\gamma}^0;\quad \hat{\gamma}^1\gamma^5 = \gamma^0 = -\gamma^5\hat{\gamma}^1;\quad \varepsilon^{\nu\sigma}\gamma_\sigma = \gamma^\nu;\quad e^a_\mu = e^\sigma \hat{e}^a_\mu;$

$e^\mu_a = E^\mu_a = e^{-\sigma}\hat{e}^\mu_a;$ When writing $\phi_\alpha = \sum_k a_\alpha(k)e^{-ikx} + a^\dagger_\alpha(k)e^{ikx} \equiv \phi^+ + \phi^-$, we say the first term has negative energy or $a_\alpha(k) = 0$ if $k_0 > 0$. Minkowski space Feynman function satisfies $\hat{\Delta} G_F = (\partial_0^2 - \partial_1^2)_x G_F(x,x') = \delta(x,x')$.

In isotropic coordinates components of curvature are:

$R^0_{101} = R^1_{001} = (\partial_0^2 - \partial_1^2)\sigma;\ R^0_{110} = R^1_{010} = (\partial_1^2 - \partial_0^2)\sigma;\ R_{0101} = e^{2\sigma}R^0_{101} = e^{2\sigma}(\partial_0^2 - \partial_1^2)\sigma;$
$R_{0110} = e^{2\sigma}R^0_{011} = e^{2\sigma}(\partial_1^2 - \partial_0^2)\sigma;\ R_{1001} = -e^{2\sigma}R^1_{001} = e^{2\sigma}(\partial_1^2 - \partial_0^2)\sigma;$
$R_{1010} = -e^{2\sigma}R^1_{010} = e^{2\sigma}(\partial_0^2 - \partial_1^2)\sigma$ ; the rest zero.
$R_{00} = R^1_{010} = (\partial_1^2 - \partial_0^2)\sigma;\ R_{11} = R^0_{101} = (\partial_0^2 - \partial_1^2)\sigma;\ R = g^{ij}R_{ij} = -e^{-2\sigma}(\partial_0^2 - \partial_1^2)\sigma$

Formulas quoted from references have made sign adjustments to be consistent with the particular notations and conventions used in this note.

APPENDIX A

CONSTRUCTION OF HILBERT SPACE FROM CAUCHY'S PROBLEM

The classical Klein Gordon equation in 1 + 1 dimensional globally hyperbolic spacetime $(M, g)$ is well posed. [13]. In this case, there is minimal coupling to geometry. This is written as

$$(\Delta + m^2)\varphi = 0 \text{ where } \Delta = |g|^{-\frac{1}{2}} \partial_\mu |g|^{\frac{1}{2}} g^{\mu\nu} \partial_\nu \tag{1A}$$

Let $S$ be the space of solutions to equation (1A) that satisfy Cauchy data at time $t=0$. Let us define

$$\pi(x) = \frac{\delta L}{\delta(\partial_0 \varphi(\underline{x}))} = |g|^{\frac{1}{2}} g^{\mu 0} \partial_\mu \varphi(\underline{x}) = |h|^{\frac{1}{2}} n^\mu \partial_\mu \varphi(\underline{x}) \quad , \tag{2A}$$

$\underline{x}$ labels a point on a surface $\Sigma_0$ of constant $x^0$ and $n^\mu$ is a unit normal to the surface. We have identified $L = \sqrt{|g|} \frac{1}{2}(g^{\mu\nu} \partial_\mu \varphi \partial_\nu \varphi - m^2)$, and $h$ is the induced metric on $\Sigma_0$. A symplectic structure on $(M,g)$ may be defined: 
$$\Omega([\varphi_1, \pi_1],[\varphi_2, \pi_2]) = \int_{\Sigma_0} (\pi_1 \varphi_2 - \pi_2 \varphi_1) dx \tag{3Aa}$$

$$= \int_{\Sigma_0} (\varphi_2 n^a \partial_a \varphi_1 - \varphi_1 n^a \partial_a \varphi_2) \sqrt{h} dx . \tag{3Ab}$$

We will now define the quantum field theory for a scalar field $\varphi$ in a basis-free way, using algebraic constructions only. The reason for this is that no unique analog of "positive energy solutions" exists in $(M,g)$. We will then need to construct a useful basis, and show how the resulting theory is unitarily equivalent under change of basis. First we need to construct a Hilbert space from the set $S$, of classical solutions to the scalar field equation.

Let us begin with the observation that along with being well-posed, unique fundamental solutions exist for $(M,g)$, which we will denote as $\Delta_A$, $\Delta_R$, called advanced and retarded Green functions. For these $(\Box + m^2)(\Delta_A f) = f$, $(\Box + m^2)(\Delta_R f) = f$; define $\Delta f = \Delta_A f - \Delta_R f$ as the Schwinger Green function which solves the homogenous Klein Gordon equation. Let us choose $\Omega$ above, and pick any bilinear map $\mu: S \times S \to \mathbb{R}$ such that for all $\psi_1 \in S$

$$\mu(\psi_1, \psi_1) = \frac{1}{4} \underset{\psi_2 \neq 0}{l.u.b.} \frac{[\Omega(\psi_1, \psi_2)]^2}{\mu(\psi_2, \psi_2)} . \tag{4A}$$

This, and all that follows up to equation (12A), is from [9].

Using Cauchy completions and quotients we can complete $S$, in the norm $2\mu$, to be a real Hilbert space $S_\mu$, with inner product $\langle \psi_1, \psi_2 \rangle = 2\mu(\psi_1, \psi_2)$. The bilinear map $\Omega: S \times S \to \mathbb{R}$ is bounded in the norm $2\mu$ so its action may be extended to $S_\mu \times S_\mu \to \mathbb{R}$ by continuity. In

Minkowski space we would now complexify $S_\mu$ and define the * operation. Here we will proceed formally and define the * operation in a general manner. First let us define the operator $J$ by

$$\Omega(\psi_1,\psi_2) = 2\mu(\psi_1, J\psi_2) = \langle \psi_1, J\psi_2 \rangle. \tag{5A}$$

From the antisymmetry of $\Omega$, it follows that $J^\dagger = -J$. From equation (19), $J$ is norm preserving in the inner product $2\mu$, and so $J^\dagger J = I$; and thus $J^2 = -I$. Therefore $J$ endows $S_\mu$ with a complex structure. Now extend the actions of $\Omega$, $J$ and $\mu$ to $S_\mu$ by complex linearity. For $\psi_1, \psi_2 \in S_\mu^{\mathbb{C}}$ define the complex inner product $\langle \psi_1, \psi_2 \rangle = 2\mu(\bar{\psi}_1, \psi_2)$. With this inner product $S_\mu^{\mathbb{C}}$ is a complex Hilbert space. Observe $iJ: S_\mu^{\mathbb{C}} \to S_\mu^{\mathbb{C}}$ is self adjoint. $S_\mu^{\mathbb{C}}$ thus decomposes into eigensubspaces of $iJ$ with eigenvalues $\pm i$, which are complex conjugates of each other. Pick $\mathcal{H}_b \subset S_\mu^{\mathbb{C}}$ corresponding to eigenvalue $+i$ to be our desired Hilbert space. In this subspace

$$\langle f, f \rangle > 0 \quad \forall f \in S_p \quad \text{and} \quad \langle f, \bar{g} \rangle = 0 \quad \forall f, g \in S_p \quad . \tag{6A}$$

For a proof, see [9]. The dual space is $\bar{\mathcal{H}}_b$. Now define a map $K: S_\mu^{\mathbb{C}} \to \mathcal{H}_b$ to be the orthogonal projection (under product of equation (6A)) onto the subspace $\mathcal{H}_b$ of $S_\mu^{\mathbb{C}}$. If $K$ is restricted to $S$, then for all $\psi_1, \psi_2 \in S$, K is a real linear map $K: S \to \mathcal{H}_b$ that satisfies

$$(K\psi_1, K\psi_2)_{S_p} = -i\Omega(\bar{K}\psi_1, K\psi_2) = \mu(\psi_1, \psi_2) - \frac{i}{2}\Omega(\psi_1, \psi_2) \quad . \tag{7A}$$

Here also
$$\operatorname{Im}(K\psi_1, K\psi_2)_{S_p} = -\frac{1}{2}\Omega(\psi_1, \psi_2). \tag{8A}$$

Now that $\mathcal{H}_b$ is defined, one may define the Fock space to be $\mathscr{F}(\mathcal{H}_b) = \bigoplus_{n=0}^{\infty} \left( \overset{n}{\otimes} \mathcal{H}_b \right),$ (9A) with the convention that $\overset{0}{\otimes} S_p = \mathbb{C}$. (24) is composed of a symmetric and an antisymmetric part

$$\mathscr{F}_s(\mathcal{H}) = \bigoplus_{n=0}^{\infty} \left( \overset{n}{\otimes}_s \mathcal{H} \right) \quad \text{and} \quad \mathscr{F}_a(\mathcal{H}) = \bigoplus_{n=0}^{\infty} \left( \overset{n}{\otimes}_a \mathcal{H} \right). \tag{10A}$$

Let $a$ and $a^\dagger$ be annihilation and creation operators on $\mathcal{F}_s(\mathcal{H}_b)$ that satisfy CCRs. For each classical observable $\Omega(\psi,\cdot) \to \hat{\Omega}(\psi,\cdot) = ia(\overline{K\psi}) - ia^\dagger(K\psi)$. For each test function $f \in C_0^\infty$, define the smeared field operator $\hat{\phi}(f): \mathcal{F}_s(\mathcal{H}_b) \to \mathcal{F}_s(\mathcal{H}_b)$ by

$$\hat{\phi}(f) \to \hat{\Omega}(\psi,\cdot) = ia(\overline{K(\Delta f)}) - ia^\dagger(K(\Delta f)) \tag{11A}$$

For a free charged (complex) scalar field, the field equations are invariant under complex conjugation and the real and imaginary parts may be treated as decoupled fields in the manner above.

The structure for a fermion field is similar and somewhat easier than the above. The real vector space $S_f$ of classical solutions (with smooth initial data of compact support) contains a natural symmetric, positive definite inner product $\Lambda: S_f \otimes S_f \to \mathbb{R}$. One can then complexify this space to obtain $S_f^\mathbb{C}$. Now complete $S_f^\mathbb{C}$ under $\Lambda$ to obtain a complex Hilbert space $S_\Lambda^\mathbb{C}$. For the one particle Hilbert space $\mathcal{H}_f$ of the quantum field theory, let $\mathcal{H}_f$ (and its conjugate space $\bar{\mathcal{H}}_f$) be any subspaces that span $S_\Lambda^\mathbb{C}$ and are orthogonal in the product $\Lambda$. The inner product of the Hilbert space is then the restriction of $\Lambda$ to $\mathcal{H}_f$. Finally, the Hilbert space of the quantum theory is the antisymmetric Fock space constructed from $\mathcal{H}_f$, $\mathcal{F}_a(\mathcal{H}_f) = \bigoplus_{n=0}^{\infty} \left( \bigotimes_a^n \mathcal{H}_f \right)$. A charged (complex) fermion field is obtained as in the Bose case above. For each $\psi \in S_f$ the operator corresponding to the classical quantity

$$\Lambda(\psi,\cdot) \to \hat{\Lambda}(\psi,\cdot) = a(\overline{K\psi}) + a^\dagger(K\psi). \tag{12A}$$

where the maps $K: S_\Lambda^\mathbb{C} \to \mathcal{H}_f$ and $\bar{K}: S_\Lambda^\mathbb{C} \to \bar{\mathcal{H}}_f$ are orthogonal projections onto the subspaces $\mathcal{H}_f$ and $\bar{\mathcal{H}}_f$ of $S_\Lambda^\mathbb{C}$, and $a$ and $a^\dagger$ are annihilation and creation operators on $\mathcal{F}_a(\mathcal{H}_f)$ that satisfy CACRs.

APPENDIX B

SCALAR FIELD FEYNMAN PROPAGATOR IN CURVED SPACETIME

Let us represent the Feynman propagator [25] as $G_{DS}^F(x,x') = i\int_0^\infty H(s,x,x')ds$ . (1B)

$H(s, x, x')$ is a function which satisfies $\left(i\dfrac{\partial}{\partial s} + \Box_x + m^2 + \xi R\right) H(s, x, x') = 0$ for $s > 0$ with the boundary condition $H(s, x, x') \to \delta(x, x')$ as $s \to 0$. Write $H(s, x, x')$ in the following way:

$$H(s, x, x') = \dfrac{1}{4\pi s} \exp\{(i/2s)[\bar{\sigma}(x, x') + i\varepsilon] - im^2 s\} \times \sum_{n=0}^{\infty} A_n(x, x')(is)^n \tag{2B}$$

The coefficients $A_n(x, x')$ are biscalar functions, symmetric in $x$ and $x'$ and regular for $x \to x'$. They are defined by the recursion relations

$$(n+1) A_{n+1} + A_{n+1;\mu} \Delta^{-1/2} \Delta^{1/2}_{;\mu} \bar{\sigma}^{;\mu} = (\Box_x + \xi R) A_n \tag{3B}$$

and the boundary condition $A_0 = \Delta^{1/2}$. Here, $2\bar{\sigma} = \sigma^{;\mu} \sigma_{;\mu}$ is the the square of the geodesic distance between $x$ and $x'$. $\Delta(x, x')$ is the Van Vleck-Morette determinant, defined by $\Delta(x, x') = -|-g(x)|^{-1/2} \det(-\bar{\sigma}_{;\mu\nu'}(x, x')|-g(x')|^{-1/2}$ and it satisfies the partial differential equation $\Box_x \bar{\sigma} = 2 - 2\Delta^{-1/2} \Delta^{1/2}_{;\mu} \bar{\sigma}^{;\mu}$ with the boundary condition $\lim_{x' \to x} \Delta(x, x') = 1$. These relations, along with (3B) are sufficient to ensure that the function $H(s, x, x')$ solves the homogeneous relation $\left(i\dfrac{\partial}{\partial s} + \Box_x + m^2 + \xi R\right) H(s, x, x') = 0$ and that the coefficients given in (3B) are unique and depend only on the geometry along the geodesic linking $x$ to $x'$. The integral (1B) can be formally solved, with the result

$$G^F_{DS}(m; x, x') = \dfrac{1}{4} \sum_{n=0}^{\infty} \dfrac{(-)^n}{(m^2)^n} A_n(x, x') \left(\dfrac{z(x, x')}{2}\right)^n H^{(2)}_{-n}(z(x, x')) \tag{4B}$$

where $z(x, x') = \left(-2m^2 [\bar{\sigma}(x, x') + i\varepsilon]\right)^{1/2}$. This may also be written as [24]

$$G^F_{DS}(m; x, x') = \dfrac{\Delta^{1/2}(x, x')}{4} \sum_{j=0}^{\infty} a_j(x, x') \left(\dfrac{-\partial}{\partial m^2}\right)^j H^{(2)}_0(\sqrt{2m^2 \bar{\sigma}}) \tag{5B}$$

with $A_0(x, x') = \Delta^{1/2}(x, x') a_0(x, x')$. Reference [25] has an extensive discussion of the development of expression (4B) as well as the different covariant representations and limits in common usage.

To put equation (4B) into the Hadamard form, let us modify (3B) to the form

$$(n+1) \bar{A}_{n+1} + \bar{A}_{n+1;\mu} \bar{\sigma}^{;\mu} + \bar{A}_{n+1;\mu} \Delta^{-1/2} \Delta^{1/2}_{;\mu} \bar{\sigma}^{;\mu} = (\Box_x + m^2 + \xi R) \bar{A}_n \tag{6B}$$

and we will have $\bar{A}_n(m; x, x') = \sum_{k=0}^{n} \frac{(-)^k}{k!} (m^2)^k A_{n-k}(x, x')$. Using $H_{-n}^{(2)}(z) = (-)^n H_n^{(2)}(z)$ one may put the propagator (and hence the two point function) in the form

$$G_{DS}^F(m; x, x') = V(m; x, y) \ln|\bar{\sigma} + i\varepsilon| + W(m; x, y) \tag{7B}$$

with $V(x, x') = \sum_{n=0}^{\infty} V_n(x, x') \sigma^n(x, x')$. $V_n$ is given by $V_n(x, x') = \frac{(-)^{n+1}}{2^n n!} \bar{A}_n$. In this formalism,

$W(x, x') = \sum_{n=0}^{\infty} W_n(x, x') \sigma^n(x, x')$ and $W_n(x, x') = \ln(m^2/2) V_n(x, x') - 2\psi(n+1) V_n(x, x')$

$-\frac{(-)^n}{2^n n!} \left[ \sum_{k=0}^{n} \frac{(-)^k m^{2k}}{k!} \left( \sum_{l=k+1}^{n} \frac{1}{l} \right) A_{n-k}(x, x') - \sum_{k=0}^{\infty} \frac{k!}{(m^2)^{k+1}} A_{n+1+k}(x, x') \right]$. Here the function

$\psi(x) = -\gamma - \sum_{l=0}^{\infty} \left( \frac{1}{z+l} - \frac{1}{1+l} \right)$ and this is defined with the $A$'s from (3B). This formalism is for the massive scalar case. Wald [9] points out that the infrared divergence of the massless case can be dealt with by setting $W_0 = 0$ and using the same recursion relations.

This line of reasoning can be applied to the $m=0$ case with (2B) replaced by

$$H(s, x, x') = \frac{1}{4\pi s} \exp\{(i/2s)[\bar{\sigma}(x, x') + i\varepsilon]\} \times \sum_{n=0}^{\infty} A_n(x, x')(is)^n \tag{8B}$$

The $A_n$ satisfy the same recursion relations as in (3B). We then write

$$G_{DS}^F(x, x') = V(x, y) \ln|\bar{\sigma} + i\varepsilon| + W(x, y). \tag{9B}$$

The integral over the $H(s, x, x')$ in equation (8B) can no longer be performed in closed form, but one nevertheless obtains an asymptotic series leading to (9B). These matters, and several others, can be found in the very readable presentation of this subject found in [27]. See also [13, 25].

APPENDIX C

FREE QUANTUM FIELDS IN CURVED SPACETIME

Let us introduce some definitions and notation. Consider a globally hyperbolic, time oriented 1+1 dimensional manifold *M* and differentiable, symmetric and non-degenerate metric *g*, denoted (*M,g*); signature $(+, -)$. The tangent bundle is denoted *TM* and the cotangent bundle is $T^*M$. For the set $\{(x,\xi) \in T^*M \mid \xi \neq 0\}$ write $T^*M \setminus \mathbf{0}$. The set of smooth sections of a vector bundle $B$ over $M$ will be denoted $\Gamma(M,B)$; sections having compact support will form the set $\Gamma_0(M,B)$.

Globally hyperbolic spacetimes admit a Cauchy surface: that is a space-like hypersurface $\Sigma \subset M$ which is intersected by any non-extendable causal curve in $M$ exactly once. For a non-compact space, the topology is essentially trivial, and we have $M = \mathbb{R} \times \Sigma$. Globally hyperbolic spacetimes are time orientable, such that the light cone, i.e. the set of all nonvanishing timelike covectors in $T_x^*M \setminus \mathbf{0}$ can be separated into a forward and backward light cone $V_x^+$ and $V_x^-$, continuously in $x$. The closed light cones $\overline{V}_x^\pm$ include the lightlike covectors. The time orientation induces the separation of the causal future $J^+(x)$ and past $J^-(x)$ of a point $x \in M$ which are the sets of all points that can be reached from $x$ by a future (past) directed causal curve.

The quantum theory of $\phi$ is defined by constructing a *-algebra of observables as follows: Consider solutions to the equation $\nabla^a \nabla_a \phi + m^2 \phi = 0$ and the algebra defined by $\hat{\phi}(F) = \int_M \hat{\phi}(x) f(x) |\det(g)|^{1/2} d^2x$. Specifically for each bounded open region $\mathcal{O} \in M$, the algebra $\mathcal{A}(\mathcal{O})$, the *-algebra generated by $\hat{\phi}(F)$ and the identity subject to the following restrictions. Let $\mathcal{D}(M)$ be the space of complex valued test functions with compact support in $M$ and $\mathcal{D}'(M)$ is the corresponding dual space of distributions; then we must have

1. For $f \in \mathcal{D}(M)$ such that $f \to \phi(f) \in \mathcal{A}(M,g)$ is complex linear;
2. $\phi((\Box_g + m^2 + \xi R)f) = 0$ for all $f \in \mathcal{D}(M)$; ($\xi = 0$ for conformal coupling in 2D)
3. $\phi(f)^* = \phi(\bar{f})$; and finally
4. $[\hat{\phi}(f_1), \hat{\phi}(f_2)] = i\Delta(f_1 \otimes f_2)I$

The algebraic notion of a state is related to the usual Hilbert-space notion by the GNS theorem. We state it here, as stated in [9]

> **Theorem**: (GNS Construction) Let $\mathcal{A}$ be a C*-algebra with identity element and let $\omega: \mathcal{A} \to \mathbb{C}$ be a state. Then there exists a Hilbert space $\mathcal{F}$, a representation $\pi: \mathcal{A} \to \mathcal{L}(\mathcal{F})$, and a vector $|\psi\rangle \in \mathcal{F}$ such that $\omega(\mathsf{A}) = \langle \psi | \pi(\mathsf{A}) | \psi \rangle$ satisfying the additional property that $|\psi\rangle$ is cyclic, i.e., the vectors $\{\pi(\mathsf{A})|\psi\rangle\}$ for all $\mathsf{A} \in \mathcal{A}$ comprise a dense subspace of $\mathcal{F}$. Furthermore the triple $(\mathcal{F}, \pi, |\psi\rangle)$ is uniquely determined (up to unitary equivalence) by these properties.

In this context a state is defined as a linear functional $\omega: \mathcal{A}(M,g) \to \mathbb{C}$ such that $\omega(1) = 1$ and positive $\omega(a^*a) > 0$ for all $a \in \mathcal{A}(M,g)$. The multi-linear functionals on $\mathcal{D}(M)$ defined by

$\omega(f_1 \otimes ... \otimes f_n) \equiv \omega(\phi(f_1)...\phi(f_n))$ are the n-point functions in this unbounded formalism. Here, a *quasi-free* state is by definition one which

satisfies $$\omega(e^{i\phi(f)}) = e^{-1/2\omega(f \otimes f)}. \qquad (1C)$$

This definition is formal. What is meant by (1C) is that set of relations obtained by functionally differentiating equation (1C) with respect to $f$.

A particular quality of quasi-free states is that all the *n*-point distributions for odd *n* vanish while the 2*n*-point distributions are made up of various permutations of products of 2 point distributions; viz: $\omega(\hat{\phi}(F_1)\hat{\phi}(F_2)\hat{\phi}(F_3)\hat{\phi}(F_4)) = \omega(\hat{\phi}(F_1)\hat{\phi}(F_2))\omega(\hat{\phi}F_3)\hat{\phi}(F_4)) +$

$$\omega(\hat{\phi}(F_1)\hat{\phi}(F_3)\hat{\phi}(F_2)\hat{\phi}(F_4)) + \omega(\hat{\phi}(F_1)\hat{\phi}(F_4))\omega(\hat{\phi}F_2)\hat{\phi}(F_3)) \qquad (2C)$$

for all test functions $F_1, F_2, F_3$ etc. The anti-commutator distribution $G(F_1, F_2) \equiv \omega(\hat{\phi}(F_1)\hat{\phi}(F_2)) + \omega(\hat{\phi}(F_2)\hat{\phi}(F_1))$ of a quasi-free state satisfies the following

(a) (*symmetry*) $G(F_1, F_2) = G(F_2, F_1)$
(b) (*weak bisolution property*) $G((\square_g + m^2 + V)F_1, F_2) = 0 = G(F_1, (\square_g + m^2 + V(x))F_2)$
(c) (*positivity*) $G(F, F) \geq 0$ and $G(F_1, F_1)^{1/2} G(F_2, F_2)^{1/2} \geq |\Delta(F_1, F_2)|$

for arbitrary space-time function $V(x)$. It can be shown [11] that, to every bilinear functional $G$ on $C_0^\infty(M)$ satisfying (a), (b) and (c) there is a quasi-free state with two-point distribution $\frac{1}{2}(G + i\Delta)$.

In curved spacetime, the existence of unitarily inequivalent representations and the inability to define a unique vacuum from any given solution necessitate a formal, algebraic approach for a rigorous definition. On the other hand, the useful Hilbert space of a given theory will be a dense subset of that provided by the above constructions, because there are additional criteria that must be included and these will place non-trivial restrictions on the set of available solutions that can be used. With the above in place, the last requirement has to do with the physical admissibility of solutions. A quasi-free state is physically admissible if (for pairs of points in sufficiently small convex neighborhoods)

(d) (*Hadamard condition*) $G(F_1, F_2) = \frac{1}{2\pi}(V(x_1, x_2) \log |\bar{\sigma}(x_1, x_2)| + W(x_1, x_2))$ ; where $2\bar{\sigma}$ is the square of the geodesic distance between $x_1$ and $x_2$ and satisfies $2\bar{\sigma} = \bar{\sigma}^{;\mu}\bar{\sigma}_{;\mu}$, and $v$ is a smooth symmetric bi-scalar function determined by the geometry and $w$ depends on the state. This is the correct generalization to curved spacetime of the known short distance behavior of the truncated two-point distributions of all physically relevant states for the case when spacetime is

flat (and *V* from criterion b vanishes). In the latter case, *V* reduces to a simple power series

$$V = \sum_{n=0}^{\infty} v_n \bar{\sigma}^n \text{ with } v_0 = \frac{m^2}{4}.$$

The Hadamard condition can be reformulated in terms of the concepts of micro-local analysis which Radzikowski [12] originally introduced as a tool towards its proof. It is then

(d') (*Wave Front Set* [or micro-local] *Spectral Condition*) WF($G+i\Delta$) = $\{(x,y;\xi,-\eta) \in T^*(M \times M \setminus \mathbf{0}) | (x,\xi) \sim (y,\eta), \xi \in \bar{V}_x^+\}$. The equivalence relation $(x,\xi) \sim (y,\eta)$ means that there is a lightlike geodesic g connecting *x* and *y*, such that the point *x* and the covector $\xi$ is tangent to $\gamma$ and $\eta$ is the vector parallel transported along the curve $\gamma$ at *y* which is again tangent to $\gamma$. On the diagonal $(x,\xi) \sim (y,\eta)$ if $\xi$ is lightlike and $\xi = \eta$.

An element $(x,\xi)$ of the cotangent bundle of a manifold (excluding the zero section $\mathbf{0}$) is in the *wave front set*, WF, of a given distribution on that manifold when the distribution is singular *at the point x in the direction* $\xi$.

The basic algebra provided by the operators constructed above does not exhaust all the observables in a theory (e.g. there are various combinations of derivatives to be properly introduced); and the difficulty of defining the product of various operators at a point in Minkowski geometry is multiplied in general spacetime. The Hadamard criteria is the appropriate method to deal with this. The stress-energy tensor is usually selected to demonstrate these issues.. A very detailed explanation of these issues in a general setting and with explicit calculations for spacetime dimensions *n* = 2 through *n* = 6 is given in [13]. This reference states that the trace anomaly for any 2D scalar field is $\frac{R}{24\pi}$. Recall further that in coordinates where the metric has the form as in (8), there is no field term coupling to *R*.

In order to define massless spinors (in 2D) the bundle of orthonormal frames (zweibeins) has to be lifted to a principal fiber bundle with the universal covering group SL(2,$\mathbb{C}$) as structure group, called the spin structure. The pull-back of the Levi-Cevita connection to the spin cover is called the spin connection, defined heuristically as $\nabla$ in equation (5). The Dirac bundle is then defined as the associated $\mathbb{C}^2$ – vector bundle. It can be shown that for any globally hyperbolic spacetime there exists such a spinor bundle *DM* [14,15].

Consider a spinor field $\psi \in \Gamma(M,DM)$ and a cospinor field $\bar{\psi} \in \Gamma(M,D^*M)$, with

$$-i\slashed{\nabla}\psi = 0, \qquad \slashed{\nabla}\psi = \gamma^\mu \nabla_\mu \psi; \tag{3C}$$

$$i\slashed{\nabla}\bar{\psi} = 0, \qquad \slashed{\nabla}\tilde{\psi} = (\nabla_\mu \bar{\psi})\gamma^\mu; \tag{4C}$$

The $\gamma$'s are the zweibein-dependent Dirac matrices and form a covariant section $\gamma \in \Gamma(M, TM \otimes DM \otimes D^*M)$. The spinor algebra is generated by solutions to (29) and (30) smeared with a smooth cospinor resp. spinor field of compact support in order to give an operator on the Hilbert space, viz:

$$\psi(v), \qquad v \in \Gamma_0(M, D^*M) \tag{5C}$$

$$\bar{\psi}(u), \qquad u \in \Gamma_0(M, DM). \tag{6C}$$

We use propagators $S_{ret}$ and $S_{adv}$ to form $S = S_{ret} - S_{adv}$ and require

$$\{\psi(v), \bar{\psi}(u)\} = iS(v, u) \tag{7C}$$

The construction of the spinor bundle leads naturally to a covariant derivative on $DM$ and the dual bundle $D^*M$ which we have denoted $\nabla$. In local coordinates one has $\nabla_\mu = \partial_\mu + \sigma_\mu$, where the 2 x 2 matrices $\sigma_\mu$ are expressed in terms of the Christoffel symbols and the Dirac matrices as shown in equation (6). The algebra $\mathcal{A}(\psi, \bar{\psi})$ is constructed similar to the above:

1. For $f \in \mathcal{D}(M)$ such that $f \to \psi(f), \bar{\psi}(f) \in \mathcal{A}(M, g)$ is complex linear;
2. $\psi((i\bar{\nabla} - m)f) = 0$ ; $\bar{\psi}((i\nabla - m)f) = 0$ for all $f \in \mathcal{D}(M)$;
3. $\bar{\psi}(f)^* = \psi(\bar{f})$; and finally
4. $\{\psi(f_1), \bar{\psi}(f_2)\} = iS(f_1 \otimes f_2)I$

The requirement of Hadamard form for the spinor two point function is a refinement of the wave-front structure called a polarization set. The two point distribution [15] $\omega^+(x, y) = \langle \Omega, \psi(x)\bar{\psi}(y)\Omega \rangle$ is a vector-valued distribution $\omega^+ \in \mathcal{D}'(M \times M, DM \otimes D^*M)$ taking values in the bi-spinor bundle $DM \otimes D^*M$. The last expression is an abuse of notation denoting the outer tensor product: the fiber over $(x, y) \in M \times M$ is $D_x M \otimes D_y^* M$ ; the first factor in the tensor product transforms as a spinor in $x$ and the second factor transforms as a cospinor in $y$. If we are given a two point function $\tilde{\omega}$ satisfying the wave equation and of Hadamard form, one extracts the Dirac two point function $\omega^+(x, y) = i\nabla\tilde{\omega}(x, y)$. Kratzert [15] states and proves the following theorem:

Let $\omega$ be a Hadamard state of the free Dirac field on a globally hyperbolic spacetime M. Then its two point function $\omega^+$ has got the following wave front and polarization sets:

$$WF(\omega^+) = \{(x, y; \xi, -\eta) \in T^*(M \times M) \setminus \mathbf{0} \,|\, (x, \xi) \sim (y, \eta), \xi \in \bar{V}_x^+\}, \tag{8C}$$

$$WF_{pol}(\omega^+) = \{(x, y; \xi, \eta, w) \in \pi^*(DM \otimes D^*M) \,|$$
$$(x, y; \xi, \eta) \in WF(\omega^+); (\mathbf{1} \otimes \mathcal{T}_\gamma(x, y))w = \lambda \cdot \xi, \lambda \in \mathbb{C}\}. \tag{9C}$$

Here $\pi$ represents the projection on the cotangent bundle and $\mathcal{T}_\gamma(x, y): D_y^*M \to D_x^*M$ denotes the parallel transport in $D^*M$ along the geodesic $\gamma$ connecting $x$ and $y$, such that $x$ is tangent to $\gamma$ in the point $x$.

## APPENDIX D

### HILBERT SPACE STRUCTURE CONDITIONS

Let V be an inner product space. A topology $\tau$ is called a *majorant* topology if (i) it is locally convex and (ii) the inner product is jointly continuous. This issue is relevant in QFT in cases where indefinite metrics are considered. In this case it often occurs that the natural topology is not jointly continuous, and it then becomes difficult to use standard methods to render V a Hilbert space.

The general utility of majorant topologies is to get a Hilbert space of states as a suitable limit of local states. A Hilbert majorant topology $\tau$ on an inner product space V makes V a prehilbert space and by a standard procedure of taking completions and quotients one obtains a Hilbert space $\mathcal{H}^\tau = \overline{V}^\tau$. The joint continuity of the inner product with respect to $\tau$ implies that the inner product can be extended to $\overline{V}^\tau$ and by the Riesz representation theorem there exists a bounded self-adjoint operator $\eta^{(\tau)}$ (called the metric) such that for $\forall \phi, \psi \in \mathcal{H}^\tau$, one has

$$\langle \varphi, \psi \rangle = \langle \varphi, \eta^{(\tau)} \psi \rangle \quad . \tag{1D}$$

Here $\langle \cdot, \cdot \rangle$ is called the Hilbert scalar product in V. The pair $(\mathcal{H}^\tau, \eta^{(\tau)})$ will be called a Hilbert space structure associated to V.

(*This section is largely a paraphrase of a section found in* [2]).